# Supervised machine learning for microbiomics: bridging the gap between current and best practices

Dudek, Natasha Katherine[a,*], Chakhvadze, Mariami[b], Kobakhidze, Saba[b,c], Kantidze, Omar[b], Gankin, Yuriy[a].

**Author affiliations**
[a] Quantori, Cambridge, USA
[b] Quantori, Tbilisi, Georgia
[c] Present address: Free University of Tbilisi, Georgia
*Corresponding author

**Abstract**

Machine learning (ML) is poised to drive innovations in clinical microbiomics, such as in disease diagnostics and prognostics. However, the successful implementation of ML in these domains necessitates the development of reproducible, interpretable models that meet the rigorous performance standards set by regulatory agencies. This study aims to identify key areas in need of improvement in current ML practices within microbiomics, with a focus on bridging the gap between existing methodologies and the requirements for clinical application. To do so, we analyze 100 peer-reviewed articles from 2021-2022. Within this corpus, datasets have a median size of 161.5 samples, with over one-third containing fewer than 100 samples, signaling a high potential for overfitting. Limited demographic data further raises concerns about generalizability and fairness, with 24% of studies omitting participants' country of residence, and attributes like race/ethnicity, education, and income rarely reported (11%, 2%, and 0%, respectively). Methodological issues are also common; for instance, for 86% of studies we could not confidently rule out test set omission and data leakage, suggesting a strong potential for inflated performance estimates across the literature. Reproducibility is also a concern, with 78% of studies abstaining from sharing their ML code publicly. Based on this analysis, we provide guidance to avoid common pitfalls that can hinder model performance, generalizability, and trustworthiness. An interactive tutorial on applying ML to microbiomics data accompanies the discussion, to help establish and reinforce best practices within the community.

**Introduction**

Machine learning (ML) is rapidly emerging as a powerful tool in biomedical research for extracting valuable insights and making accurate predictions from vast and complex datasets (MacEachern & Forkert, 2021; Shehab et al., 2022). The microbiome field is no exception, with ML showing great potential for innovations in disease screening, diagnostics, prognostics, biomarker discovery, and therapeutic development (Curry, Nute, & Treangen, 2021; Marcos-Zambrano et al., 2021; McCoubrey, Elbadawi, Orlu, Gaisford, & Basit, 2021; McCoubrey, Gaisford, Orlu, & Basit, 2022), with the emergence of clinical ML-based microbiome tools for the commercial space having been introduced by companies such as Karius, Viome, Micronoma, BiomeOne, and others.

Common technical challenges of applying ML to microbiome datasets include small dataset sizes (Moreno-Indias et al., 2021; Navas-Molina, Hyde, Sanders, & Knight, 2017; Papoutsoglou et al., 2023), the integration of multiple data modalities (Daliri, Ofosu, Chelliah, Lee, & Oh, 2021; Graw et al., 2021; Navas-Molina et al., 2017), and the subtleties of standard ML evaluation paradigms (Thomas P. Quinn, n.d.; Wiens et al., 2019). Furthermore, successful projects often require teams with expertise in bioinformatics, ML, and the domain of study (e.g., deep medical understanding of a specific disease) (Wiens et al., 2019). These challenges and shortfalls in basic ML standards as applied to microbiomics can contribute to ambiguity and imprecision of the ML studies, which ultimately is a pitfall in microbiome research.

In order to identify common strengths and weaknesses in research applying ML to microbiomics, one can quantify and evaluate the occurrence of selected experimental design parameters (e.g., type of feature being used for ML). This data-driven approach allows for the promotion of best practices and the correction of commonly encountered shortcomings. Here, we present a framework for designing ML experiments in the microbiome space. First, we aim to characterize broad trends in the application of ML to microbiome studies, focusing on aspects such as dataset size, features utilized for ML, model evaluation methods, data leakage, and reproducibility. This is achieved through a review of 100 journal articles in which authors applied supervised ML to human microbiomics data. Our review is intentionally broad in order to inform practitioners about the diversity of methodologies and available approaches and patterns in their application. Second, insights from current practices are utilized to highlight key challenges faced by ML practitioners in this field and we discuss potential strategies for addressing them. Third, we provide an accompanying interactive online tutorial which serves as a hands-on demonstration of best practices for applied ML for the microbiome field. Adherence to the best practices outlined here will improve model performance, enhance reproducibility, and advance the overall quality of ML applications in the microbiome space.

**Part 1: Charactering current practices in the application of ML to microbiomics data**

The following section provides a characterization of current practices in the application of ML to microbiomics data (also available in short form in **Supplementary Table 1**). We quantify attributes such as the types of tasks being performed using ML, dataset characteristics, and the types of features being generated, as well as the methodology used to train and evaluate model performance.

*Survey of the microbiomics-ML literature*

To identify common practices and recurring problems in the application of ML to microbiomics data, we performed a review and analysis of the literature **(Figure 1)**. To identify relevant journal articles, on March 1, 2023 we searched PubMed for open-access journal articles published in 2021-2022 whose title and/or abstract included the terms "microbiome"/"microbiota" and "machine learning". These search terms were intentionally selected to avoid biasing the search to any one sequencing methodology or experimental approach (e.g., searching for "metatranscriptomics" would have biased our search towards studies using this type of data). We then screened papers for the following inclusion criteria: 1) research articles reporting the results of an experiment, 2) articles that apply supervised ML, 3) articles that use 'omics data derived from the microbiome in their final ML model, 4) articles that use data from the human microbiome to study a question related to human health, 5) articles that present results from an applied scientific task – articles on engineering novel frameworks, tools, or applications were excluded. Of 196 journal articles identified by the original PubMed search, the most recent 100 articles that met these criteria were retained for the literature review (see Supplementary Data 1 for the full list of articles) **(Figure 1A)**. This sample size (n=100) was chosen to strike a balance that allows us to provide a high-level overview within our resource constraints; conducting an in-depth review of each study's methodology by multiple individuals (see below) is a time-consuming endeavor. Articles spanned 57 journals, with a median impact factor (2022) of 6.073 and a range of 0.602 to 87.244 **(Figure 1B)**; 97% had an impact factor >1. One journal had not yet been assigned an impact factor due to its recent establishment. The most frequently encountered journals were Frontiers in Microbiology (n=11), Microbiology Spectrum (n=6), and Frontiers in Cellular and Infection Microbiology (n=6), Gut Microbes (n=5), and Scientific Reports (n=5) **(Figure 1C)**. For multi-task studies, our analysis was conducted on the first task presented. This choice of observation criteria reflects the observation that a single group of authors typically employed a similar approach across multiple tasks. For example, a study may have attempted to predict presence/absence of diseases A, B, and C from the microbiota, each as a unique task rather than as a multi-class classification problem. Counting each task multiple times would have resulted in some studies being overrepresented in our survey. Instead, we opted to sample the first task presented in each study, thereby ensuring a balanced representation of the methods used by each author group.

For each study, ML experimental design was analyzed by two reviewers: a data scientist with domain expertise in microbiomics and one of two ML engineers. Discrepancies amongst our reviewers in the interpretation of specific points in manuscripts were discussed until resolved or designated as information having been "not recovered" from the manuscript. A final review of all data collected was performed by the domain expert data scientist to ensure consistency.

*Types of ML tasks*

The optimal ML approach for a task will depend on the domain and nature of the problem. The five most common medical domains observed in our corpus were oncology (n=26), gastroenterology (n=25), pediatrics (n=17), infectious disease (n=14), and hepatology (n=12) **(Figure 2A)**. ML tasks could be broadly classified as related to disease or disease subtype diagnosis (n=69), prognosis of disease progression or response to treatment (n=22), or predicting a host characteristic (e.g., age, diet) (n=9). Tasks were framed as classification (n=90) or regression (n=10) problems, with classification problems largely being binary (n=85) but including tasks with up to 9 classes. Biomarker discovery was widely cited as a motivating factor for applied ML research, with 62 studies actively performing interpretability analyses to identify features contributing to model performance.

*Dataset characteristics*

The quality and utility of ML models is fundamentally limited by the quality of the data on which they have been developed. Inappropriately small datasets have been cited as one of the major shortcomings of studies applying ML to microbiomics data (Moreno-Indias et al., 2021). For the 86 studies for which we were able to recover the dataset size used for ML tasks, the median microbiomics dataset size was 161.5 samples and ranged from 10 - 34,057 samples per study **(Figure 2B)**. Notably, for 14% of the articles we were unable to recover the size of the dataset. Here, 73% of studies had <1000 samples and over a third had <100 samples (n=31). The largest dataset surveyed (n_samples=34,057) was privately owned by a direct-to-consumer microbiomics company. Neural networks, in particular, often require large amounts of training data. Of five studies observed here to apply neural networks, four reported dataset size, with sample sizes ranging from 79 - 3,595 samples. In some cases, data was fully sourced from

(n=28) or supplemented with (n=4) data from external sources. Next we evaluated the relative size of target classes in datasets used for classification problems, as when one or more classes are underrepresented compared to others, it can introduce bias and harm model performance ("Sampling and Splitting," n.d.). Applying Google's recommended thresholds for categorizing the severity of dataset imbalances ("Sampling and Splitting," n.d.), datasets were found to be balanced (n=29), mildly imbalanced (n=15), or moderately imbalanced (n=9) **(Figure 2C)**. We identified no severely imbalanced datasets.

Demographic factors have been associated with distinctive microbial signatures (Chen et al., 2016; Gacesa et al., 2022; Kaplan et al., 2020; Scepanovic et al., 2019), suggesting that predictive algorithms learned for one population may not transfer well to others. For instance, the microbiota undergoes significant changes throughout an individual's lifespan, with distinct microbial community compositions observed in infants compared to seniors (Badal et al., 2020). Therefore a model trained in the infant microbiota would be unlikely to generalize well to that of seniors. Here, the breadth of demographic data used to describe populations varied. The country in which subjects were living, or at least receiving medical care, was reported in 76 out of 100 studies, with cohorts deriving from Asia (n=42), Europe (n=26), North America (n=21), South America (n=3), Oceania (n=2), and Africa (n=2) **(Figure 2D)**. Meanwhile, 72 studies reported cohort summary statistics about age, 66 reported cohort gender composition, 11 reported cohort race or ethnicity composition, two reported cohort education levels, and none reported cohort income levels **(Figure 2E)**. A word search for the term "bias" in the text of manuscripts suggested that seven papers directly addressed this topic.

*Generating and selecting features for ML*

The choice of ML feature will have an impact on the predictive power of a model. In the literature, types of 'omics data used to study the human microbiome consisted of metagenomic (n=91), metabolomic (n=10), metatranscriptomic (n=5), metaproteomic (n=1), with 7/100 studies utilizing two microbiome data modalities for ML **(Figure 2F)**. Of these seven studies, five integrated data from both modalities to develop a model, while the remaining two studies built independent models for each data modality. An additional seven studies integrated a form of microbiome data with human-derived 'omics data for model development, namely metabolomic (n=3), transcriptomic (n=2), genetic (n=1), or radiomic (n=1) data.

Within metagenomics, numerous feature options were possible (**Supplementary Table 2**), with most studies inferring taxonomic composition (n=94). Of these 94 studies, most performed their analysis at the genus (n=38) or species level (n=32). Such data is compositional, which must be accounted for when generating features (G. B. Gloor & Reid, 2016; Gregory B. Gloor, Jean M. Macklaim, Vera Pawlosky-Glahn, Juan J. Egozcue, 2017a; Tsilimigras & Fodor, 2016). Of 94 studies that used taxa as features, we estimate that 33-65 used relative abundance as a transformation prior to ML. Feature transformation was often not explicitly stated, but relative abundance is likely given that it is the default generated by many tools. Alternatives consisted of a) the centered log-ratio transform (n=11), b) other forms of log transform such as log2 (n=9), c) rarefied counts (n=3), d) presence-absence (n=3), and e) cum sum normalization (n=1).

*Training an ML model*

Training a model consists of teaching an ML algorithm to recognize patterns and make predictions or decisions based on data. Design choices will impact the patterns a model learns and its capacity for generalization.

Before training, data must be split into partitions (e.g., training, test, and validation datasets) so that model performance can be evaluated on previously unseen data. Studies implemented varied forms of experimental design, including static data partitions, repeated k-fold cross-validation (including leave-one-out cross-validation), Monte Carlo cross-validation, leave-one-dataset/cohort/family-out cross-validation, and nested cross-validation. Methodology was often not easily interpreted, obfuscating analysis. Most studies (n=58) implemented an experimental design that included training and evaluating model performance across repeated partitionings of the data (e.g., multiple training-test splits), while 26 appeared to implement a single, static data partition **(Figure 3A)**. Of the remaining 16 studies, we did not recognize any data partition in nine and were unable to recover methodology in another seven **(Figure 3A)**. We did not observe a relationship between journal impact factor and whether a static vs repeated data partition was used (Mann-Whitney U test, one-sided, p=0.55).

Feature selection is the process of identifying and selecting the most relevant features or variables from a dataset and is often applied to high-dimensional datasets in an effort to reduce overfitting. Here, 61 studies reviewed performed feature selection, 31 did not, and for eight we were unable to recover this information. Feature selection was performed using filter methods (e.g., univariate feature selection, DESeq2, LEfSe), wrapper methods (e.g., recursive feature elimination), or embedded methods (e.g., random forest, L1 logistic regression). It should be noted that L1, and not L2, logistic regression can be applied for feature selection.

There is no single algorithm that will outperform every other algorithm on all tasks (Wolpert & Macready, 1997). Here, we found that 44 out of 97 papers compared the performance of at least two learning algorithms **(Figure 3B)**. The number of algorithms compared did not correlate with the impact factor of the journal in which a study was published (Spearman's rank correlation, p=0.17). Of 93 studies for which we were able to identify the final model, the top five learning algorithms most commonly applied for training the final model consisted of random forest (n=47), logistic regression (n=10), XGBoost (n=9), Support Vector Machines (SVMs) (n=7), and neural networks (n=5) **(Figure 3C)**. We observed that of 52 studies that applied a single learning algorithm, 62% (n=33) applied random forest by default. Notably, of those that did compare the performance of random forest against at least one other learning algorithm (n=47), 30% (n=14) selected a learning algorithm other than random forest. We did not observe a relationship between the final learning algorithm selected as a function of the impact factor in which a study was published (Kruskal-Wallis, p=0.21).

Hyperparameter tuning can have a strong impact on model performance (Olson, Cava, Mustahsan, Varik, & Moore, 2018; Sanders & Giraud-Carrier, 2017). Here, we identified 38 out of 100 studies that reported performing hyperparameter tuning **(Figure 3D)**. The final hyperparameters used for training were identified, at least partially, in 29 studies, while search sweeps were identified in 10 studies **(Figure 3D)**. We did not observe a relationship between the impact factor of the journal in which a study was published and whether they reported having performed hyperparameter tuning (Mann-Whitney U test, one-sided, p=0.14), reported the final hyperparameters (Mann-Whitney U test, one-sided, p=0.29), or reported hyperparameter search sweeps (Mann-Whitney U test, one-sided, p=0.71).

Publicly sharing code used to train a model aids methodological transparency and reproducibility (Pineau et al., 2021). Here, 22 studies provided a functional link to code (even if only partial) that was ostensibly used to implement experiments described in the paper **(Figure 3D)**. Another three studies stated or implied that code was available upon request. We did not observe a relationship between the impact factor of the journal in which a study was published and whether the study made code publicly available (Mann-Whitney U test, one-sided, p=0.35). Even with code, to reproduce non-deterministic processes one must set a random state. Here, we identified 18 studies that reported setting a seed for any non-deterministic process during the development of their ML model/s **(Figure 3D)**. We did not observe a relationship between the impact factor of the journal in which a study was published and whether the study reported having used a seed for any non-deterministic process (Mann-Whitney U test, one-sided, p=0.09).

*Evaluating model performance*

Effective evaluation of model performance allows one to gauge its suitability for real-world applications. Most studies here (n=58) performed replicates of their ML experiment across multiple data partitions (e.g., training-test splits). Of 57 studies for which we recovered an estimate of model performance, 58% (n=33) reported an evaluation metric/s as measures of central tendency (i.e., mean, median, or unspecified which) without providing summary statistics to characterize the distribution of results. Of the 74 studies that compared the performance of two or more models, 15% (n=11) performed statistical testing to support at least some claims that one or more models had superior performance **(Figure 3E)**. This was most commonly achieved using the Wilcoxon rank-sum test (n=3). We did not observe a relationship between the impact factor of the journal in which a study was published and whether a study compared two or more models (Mann-Whitney U test, one-sided, p=0.71) or used statistical testing to support claims of model superiority (Mann-Whitney U test, one-sided, p=0.15). While the relative infrequency of rigorous statistical analysis is consistent with what was observed in ML at large, it is a surprising divergence from the practices of most scientific and engineering communities (Henderson et al., 2018; Pineau et al., 2021).

After selecting the final model, one must evaluate how its performance compares to that of other methods for accomplishing the given task. In reviewing the literature, 10 out of 100 studies explicitly stated that they compared performance against baseline or benchmark models. While admittedly somewhat subjective, we inferred that 13 studies employed baseline models and 14 studies employed benchmark models **(Figure 3E)**. We did not observe a relationship between the impact factor of the journal in which a study was published and whether a study employed a benchmark model (Mann-Whitney U test, one-sided, p=0.72) or baseline model (Mann-Whitney U test, one-sided, p=0.43).

Beyond aggregate measures of performance, evaluating the distribution of errors amongst different cohorts in the dataset is necessary for developing models that are reliable, fair, and trustworthy. Here, we loosely defined error analysis as an attempt to understand the distribution of errors in the final model beyond aggregate class-level performance metrics. We identified four studies that did so **(Figure 3E)**. We did not observe a relationship between the impact factor of the journal in which a study was published and whether a study performed error analysis (Mann-Whitney U test, one-sided, p=0.48).

Interpretability analyses provide insights into how models make decisions(Lundberg & Lee, 2017). In the literature, 62 out of 100 studies performed interpretability analyses **(Figure 3E)**. While methodology was often not described, common approaches included analysis of SHAP values, Gini impurity decrease index, and permutation feature importance. We did not observe a relationship between the impact factor of the journal in which a study was published and whether a study evaluated feature importance (Mann-Whitney U test, one-sided, p=0.23).

*Model verification failure*

Evaluating a model on previously unseen data enables the estimation of its performance on new, real-world data. Failure modes that can arise when validating models include test set omission and test set leakage (T. P. Quinn, 2021). The former arises when the final model's performance is not validated on a test set that is independent from the training dataset, while the latter arises when information from the test set is used during training. In both cases, the reported performance of a model is likely to be overestimated (T. P. Quinn, 2021; Varma & Simon, 2006). In reviewing the literature, we assessed the extent to which there was conclusive evidence that no test set omission or test set leakage had occurred. Our team was confident in the robustness of model verification procedures in only 14 out of 100 cases.

The most common three issues encountered with model verification were **(Figure 3F)**:
1. Model assessment and selection was performed directly on the test set used for hyper-parameter tuning, making the final performance estimate unreliable as a predictor of performance on new data. Despite this, the final model's performance on the test set was reported as an estimate of how well the final model would perform on previously unseen data (n=62).
2. Feature selection appeared to have been performed on the entire dataset, rather than on the training set/s alone (n=47).
3. No training-test split appeared to have been implemented and thus there was no test set (n=13).

We next assessed whether there was a difference in the impact factor of journals that published articles with vs without potential model validation failure, via the application of a one-sided Mann

Whitney U test. We found no difference (p=0.25), suggesting that this issue is widespread across journals.

**Part 2: Bridging the gap between current and best practices in the application of ML to microbiomics data**

Based on our review of current practices, we identified 10 primary recurring shortcomings with how ML is applied to microbiomics data. Discussion of challenges and solutions is provided **(Figure 4)**.

1. **Carefully identify and frame appropriate ML tasks**

   A good task will be scientifically and/or clinically impactful, realistic and achievable given available resources and technologies, and follow a user-centric design where applicable (Beede et al., 2020; Wiens et al., 2019). Task intent should be clearly conveyed to stakeholders by including description of specific target labels (e.g., survival time in months) and features, rather than via vague descriptors such as "patient stratification" and "microbiome features". Biological problems must be translated into an optimal form for ML and the justification for design choices must be conveyed to readers. For example, we noted that in the literature continuous target variables that would have been amenable to regression tasks were often discretized into classes for classification without justification for the chosen approach. When overly broad discretization bins are applied, the model's performance may be adversely affected. Therefore, this design choice warrants careful consideration by ML practitioners and should be thoroughly justified to readers.

   For further discussion on how to identify and frame machine learning tasks, we refer the reader to (Wiens et al. 2019; Scott et al. 2021; Ghassemi et al. 2020; Kelly et al. 2019).

2. **Assess project feasibility given available resources**

   While ML successes have rightfully gained significant attention, it is important to acknowledge that not all tasks are suitable for an ML approach. A primary concern in the microbiomics field is the limited quantity of data available.

   For any given task, the minimum amount of data required will depend on factors such as the structure of the underlying data and the complexity of the task. A heuristic commonly applied in the field of ML is that there should be ≥10X more instances than there are features (Lakshmanan, Robinson, & Munn, 2020). In some cases, identifying the dataset size used for similar tasks in the literature may provide additional insights. If preliminary data is available, learning curves plotting the performance of a simple model (y-axis) vs how much data it was trained on (x-axis) can provide insights into whether more data is likely to be beneficial.

Other feasibility considerations include required vs available data quality (e.g., target label consistency), team composition (i.e., clinical, bioinformatic, and ML expertise) and computational resources.

For a detailed discussion on assessing the feasibility of machine learning projects, we direct the reader to (Scott et al. 2021; Verma et al. 2021; Rajput et al. 2023; L'heureux et al. 2017; Amershi et al. 2019).

3. **Mitigate bias and enhance fairness**

   Diverse demographic factors have been associated with distinct microbial signatures (Chen et al., 2016; Gacesa et al., 2022; Kaplan et al., 2020; Scepanovic et al., 2019). For example, the vaginal microbiota of healthy women living in North America differs by ethnic background (Ravel et al., 2011) and associations between the vaginal microbiota and the risk of preterm birth vary between study cohorts of different racial compositions (Callahan et al., 2017). This is of concern because models trained on one patient population may not generalize well to others.

   Fair and equal access to high quality medical necessitates tools that work well for patients from different demographic backgrounds. To develop models with improved generalizability, one can implement methods such as data balancing (Parmar et al. 2023; de la Cruz-Ruiz et al. 2024; Yang and Zou 2021), robust cross-validation (e.g., with stratified sampling) (Szeghalmy and Fazekas 2023; Yates et al. 2023), bias detection (Alelyani 2021; Pagano et al. 2022), data drift detection (Chi et al. 2022), and the use of fairness-aware algorithms, libraries, and frameworks (Tizpaz-Niari et al. 2022; Iosifidis et al. 2019; Pessach and Shmueli 2022).. Furthermore, investigators should collect and report study demographics so that readers are informed as to the population/s studied and can assess the model's applicability to other groups or contexts (Bozkurt et al. 2020; Sahiner et al. 2023).

   For further discussion of bias and algorithmic fairness, we refer the reader to (McCradden et al. 2020; Sirugo et al. 2019; Obermeyer et al. 2019; Zou and Schiebinger 2018; Futoma et al. 2020).

4. **Account for the compositional nature of microbiomics data**

   Compositionality of microbiome data is often addressed by converting read counts per sample to relative abundances. Doing so, however, can lead to spurious correlations between features that may hinder model performance (G. B. Gloor & Reid, 2016; G. Gloor, Macklaim, Pawlowsky-Glahn, & Egozcue, 2017; Tsilimigras & Fodor, 2016). Log ratio transformations, such as the centered or isometric log-ratio transformations, offer an attractive approach to working with compositional data (Gregory B. Gloor, Jean M. Macklaim, Vera Pawlosky-Glahn, Juan J. Egozcue, 2017a, 2017b).

We refer the reader to (Gregory B. Gloor, Jean M. Macklaim, Vera Pawlosky-Glahn, Juan J. Egozcue 2017b, 2017a) for deeper discussion of how to account for compositional data.

5. **Understand and avoid test set omission and leakage**

   Flawed estimation of model performance is widespread in the 'omics literature (T. P. Quinn, 2021; Teschendorff, 2019; Whalen, Schreiber, Noble, & Pollard, 2022). Here, we could not confidently rule out test set omission or leakage in 85% of studies. This is in line with Quinn 2021's (T. P. Quinn, 2021) estimate that the conclusions of 88% of microbiomics ML papers cannot be trusted at face value due to improper model verification processes. This casts serious doubt on the reported predictive power of many microbiome-based ML models.

   While there are diverse ways by which model verification failure can occur, three guidelines to avoid the most common mistakes are to 1) ensure that a dedicated test set is allocated for evaluation purposes, whether as part of a static or repeated split design, 2) perform feature selection exclusively on the training data and not on the entire dataset, and 3) avoid performing hyperparameter tuning or model selection on the same data partition used for model evaluation,

   For additional insights into what constitutes model verification failure and how to avoid it, we refer the reader to (Teschendorff 2019; Whalen et al. 2022).

6. **Consider using repeated k-fold or nested cross-validation to train and evaluate models when dataset size is limited**

   When a dataset and therefore test set is small, the variance uncertainty of performance estimates increases (Beleites, Neugebauer, Bocklitz, Krafft, & Popp, 2013). Cross-validation schemes such as repeated k-fold cross-validation and nested cross-validation (**Figure 5**) can improve estimates of performance in such cases - when correctly applied - because they allow multiple experimental replicates to be run using different training-test splits. Summary statistics calculated across replicates provide a more accurate estimate of the model's performance.

   A demonstration of how to apply repeated k-fold cross-validation to microbiomics data, without test set omission or leakage, is included in the accompanying interactive tutorial.

   For further discussion of cross-validation, we refer readers to (Topçuoğlu et al. 2020; Yates et al. 2023).

7. **Perform a thorough exploration of ML experimental variables during training**

Default strategies and settings are selected to give good performance on a wide range of tasks; they are not designed to deliver the best performance on a specific task of interest. Thorough exploration of experimental variables such as learning algorithms (Wolpert & Macready, 1997) and hyperparameters (Olson et al., 2018; Sanders & Giraud-Carrier, 2017) are a relatively low cost addition to experimental design that can lead to major gains in model performance. For example, Olson et al., 2018 (Olson et al., 2018) tested the performance of 13 supervised classification algorithms on 165 datasets, with and without hyperparameter tuning. They found that while HP tuning usually resulted in modest improvements to model performance (3-5%), it had the potential to produce improvements of up to 50%.

For further discussion of exploring experimental parameters and hyperparameters, we refer the reader to (Wolpert and Macready 1997; Hutter et al. 2019)

8. **Comprehensively evaluate model performance**

The results of a machine learning model typically require contextualization by a domain expert in the field of application. This includes identifying and mitigating the potentially deleterious impact of confounding factors which could harm model performance and results interpretation. No single evaluation metric is adequate for assessing all essential characteristics of a model (Hicks et al. 2022; Park and Han 2018; Reyna et al. 2022). For example, when evaluating binary classification models, the Area Under the Receiver Operating Characteristic Curve score (AUROC) offers a broad view of model discrimination ability across various thresholds, while the F1 score provides a focused assessment of the model's performance in terms of precision and recall, making each metric valuable for different contexts and objectives. Including multiple evaluation metrics such as true positives, false positives, true negatives, false negatives, AUROC, and F1 scores provides a more holistic understanding of model performance (Hicks et al. 2022; Park and Han 2018; Reyna et al. 2022).

When implementing experimental replicates – which is typically recommended when working with small datasets common to microbiomics – performance of the model may vary due to random factors such as the exact split and the initialization of the model (Topçuoğlu et al., 2020). Thus when comparisons are made between multiple models, they should be validated through statistical testing. Similarly, if one is benchmarking the predictive value of microbiomic data against traditional risk factors, one must corroborate that a perceived difference is not simply due to random chance (e.g., sampling error). Notably, baseline and benchmark models provide important context for evaluating the predictive power of models and should be routinely incorporated into experimental design.

Post-hoc interpretability analyses, such as SHAP analysis, provide further valuable insights into how and why a model is making predictions (Lundberg & Lee, 2017). Meanwhile, error analyses provide insights into important demographic factors that affect model performance and can help identify potential failure modes to improve upon in future works (Nushi, 2021).

For further reading on various methods for model evaluation, we refer the reader to (Hicks et al. 2022; Park and Han 2018; Reyna et al. 2022). For additional discussion of post-hoc interpretability analyses we suggest (Zednik and Boelsen 2022; Reddy 2022). For discussion of error analysis we refer the reader to (Nushi 2021).

9. **Ensure that your work is reproducible.**

   Reproducibility has been widely cited as a barrier to the advancement of science (Beam, Manrai, & Ghassemi, 2020; Heil et al., 2021; Pineau et al., 2021). Here, we noted that key information about experimental methodology was often missing from the articles in our corpus, including complete descriptions of features used for ML, ML experimental design (e.g., nested vs repeated k-fold cross-validation), hyperparameter search sweeps and final settings, and seeds set for non-deterministic processes.

   Sharing research code enhances transparency and reproducibility (Cadwallader, Mac Gabhann, Papin, & Pitzer, 2022). Here, we found that 78% of papers refrained from publicly sharing their ML code. Even within the biomedical field, where 50.1% of studies do not share code (Sharma et al., 2024), this figure is concerning.

   Techniques for enhancing reproducibility include using version-controlled repositories, sharing preprocessed datasets, documenting computational environments, applying open data and code sharing, performing random seed control, and providing clear documentation of code, amongst others. Figures demonstrating experimental design are also a powerful method for improving methodological transparency.

   To combat inadequacies in reporting, which harms reproducibility, numerous reporting checklists have been developed for both study authors and reviewers. Biomedical study checklists include STORMS (Mirzayi et al. 2021), which is a microbiome-specific checklist, and clinical trial reporting checklists such as TRIPOD (Collins et al. 2015; Heus et al. 2019) and SPIRIT (Chan, Tetzlaff, Gøtzsche, et al. 2013; Chan, Tetzlaff, Altman, et al. 2013). In the ML domain, resources include the NeurIPS Paper Checklist ("NeurIPS 2021 Paper Checklist Guidelines" 2021) and the Papers With Code ML Code Completeness Checklist (Stojnic 2020). Some checklists span both ML and life sciences or healthcare, such as PROBAST (Wolff et al. 2019).

   Furthermore, the adoption of workflow management systems can facilitate the creation of reproducible data analysis pipelines. These tools are designed to automate and streamline the execution of complex workflows, ensuring that every step is documented and consistently executed. Meanwhile various frameworks have been developed to help manage the machine learning lifecycle, providing functionalities for experiment tracking, version control, and model deployment.

For comprehensive discussion on reproducibility standards in ML, we refer the reader to (Pineau et al. 2021; Beam, Manrai, and Ghassemi 2020; Heil et al. 2021), in addition to the tools and reporting checklists described above.

**10. Keep abreast of the latest research developments and explore emerging techniques**

ML innovations are poised to tackle key challenges in the microbiome field. For example, small dataset sizes may be partially mitigated through the adoption of synthetic data augmentation techniques specifically for microbiomics data (Díez López, Montiel González, Vidaki, & Kayser, 2022; Gordon-Rodriguez, Quinn, & Cunningham, n.d.; Mulenga, Kareem, Sabri, & Seera, 2021; Reiman & Dai, 2020; Sayyari, Kawas, & Mirarab, 2019) or novel embedding techniques for microbiome features that reduce the dimensionality of these datasets (Woloszynek, Zhao, Chen, & Rosen, 2019). Fine-tuning of microbiome foundation models may one day become the norm, akin to work done on developing a foundation model for single-cell multi-omics data (Cui, Wang, Maan, & Wang, 2023). Meanwhile, advancements in the integration and analysis of multi-omics data, including microbiomics data, will yield deeper insights into the dynamic interactions among various omics layers and allow for more accurate modeling of biological systems (Arıkan and Muth 2023; Daliri et al. 2021).

Not all innovations will be unambiguously positive; some may expose vulnerabilities that could be exploited by malicious actors. For example, emerging evidence suggests training datasets can be reconstructed from the parameters of neural networks and other ML models (Balle, Cherubin, & Hayes, 2022; Haim, Vardi, Yehudai, Shamir, & Irani, n.d.; Wang & Kurz, 2022), potentially compromising data governance and patient privacy.

Microbiomics practitioners can stay up to date on innovations in ML by attending ML conferences (e.g., the Conference on Neural Information Processing Systems (NeurIPS), the International Conference on Machine Learning (ICML), and the International Conference on Learning Representations (ICLR)), regularly reviewing new scientific publications (e.g., by setting up alerts to receive notifications on relevant research developments), and subscribing to newsletters from leading research institutions, amongst other methods.

**Part 3: Tutorial the application of ML to microbiomics data**

To encourage the adoption of best practices in constructing ML models for microbiome research, we developed an interactive tutorial that guides users through the development, evaluation, and interpretation of ML models for predicting whether an individual has schizophrenia based on the composition of their fecal microbiota. The goal of the tutorial is to evaluate: a) whether the microbiome has predictive value for the diagnosis of schizophrenia, and if so, b) which methodological ML parameters (i.e., learning algorithm, features) yield the highest predictive results, and c) which microbial taxa provide value for predicting whether an individual has schizophrenia. For this purpose, shotgun sequencing data was obtained for 171 fecal samples; 90 from medication-free individuals with diagnosed schizophrenia and 81 from individuals without diagnosed schizophrenia (Zhu et al., 2020). Repeated 10-fold CV was implemented to perform hyperparameter tuning and model training without data leakage. We applied SHAP to identify microbial taxa with the highest predictive value for schizophrenia and error analysis to gain insights into the distribution of errors across subsets of the study population. At each step, we provide detailed explanations of the rationale behind experimental design decisions, offer guidance on the application of best practices, and discuss the interpretation of the results. The tutorial is available at https://github.com/quantori/tutorial-on-ML-for-microbiomics.

**Discussion**

Microbiomics holds great potential for clinical applications, with ML set to transform our ability to gain meaningful insights from and harness the full utility of such datasets. Here, we employed a data-driven approach to characterize current practices in the application of ML to microbiomics data, identify where they diverge from best practices, and provide discussion on how to bridge this gap. To further support the application of ML best practices, we developed a tutorial tailored to the microbiomics community that provides an interactive demonstration of how to apply best practices when developing an ML model for microbiomics data.

Through our analysis of current practices in the application of ML to microbiomics, several issues emerged as field-wide challenges that will likely require concerted effort to overcome. The first is that the size of microbiome datasets tends to be relatively small for ML, likely resulting in high bias, high variance, and diminished model performance (Lakshmanan et al., 2020). In the present sample of the literature, the median dataset size was 161.5 samples, with 73% of studies having <1000 samples and over a third having <100 samples. In ML, the "rule of 10" approximates that dataset size is ideally on the order of ten times as many features as are present (Lakshmanan et al., 2020). If we use our tutorial on ML for microbiomics as an example, the final dataset which used microbial species as features consisted of 1,082 features. This suggests that the ideal dataset size would be approximately 10,820 samples; 171 were available. It is essential that ML practitioners in the microbiomics field be aware of the deleterious effects of small dataset size and account for it during experimental design, model verification, and results interpretation (Vabalas, Gowen, Poliakoff, & Casson, 2019; Ying, 2019). In the long term, data accumulation in repositories managed by large consortia will play a pivotal role in future efforts to source large datasets to power data-hungry ML algorithms. This data must be managed such that it is Findable, Accessible, Interoperable, and Reusable (FAIR) (Moreno-Indias et al., 2021). This responsibility extends beyond consortia; individual researchers must also contribute by submitting detailed and clear metadata alongside their data, and such behavior should be incentivized. Furthermore, the field must establish robust frameworks for developing and evaluating models that integrate data from multiple sources, where confounding variables are prevalent.

Another striking observation that resulted from our analysis was that important demographic data was often limited or missing in the literature reviewed here. This is problematic because characterizing the demographics of study participants provides context required to interpret study findings accurately, assess their relevance to broader populations, and ensure that research is conducted fairly and ethically (Huang, Galal, Etemadi, & Vaidyanathan, 2022; Perez-Downes et al., 2024; Starke, De Clercq, & Elger, 2021; Vokinger, Feuerriegel, & Kesselheim, 2021; Wiens et al., 2019). Here, we found that the most commonly reported demographic features were country of residence, age, and sex (76%, 72% and 66% of studies, respectively). Additional demographic attributes such as race/ethnicity, education level, and income were rarely, if ever, reported (11%, 2%, and 0% of studies, respectively). Such demographic factors are known to influence microbiome composition (Callahan et al., 2017; Chen et al., 2016; Kaplan et al., 2020; Ravel et al., 2011; Scepanovic et al., 2019), and thus have the potential to introduce bias into ML models. This could lead to models that perform well with one demographic but fail with others, posing a significant challenge when attempting to implement these models in clinical settings. For instance, we observed that studies involving individuals from the least studied continents (tied: Africa and Oceania) were 21 times less prevalent than those focusing on the most extensively studied continent (Asia). This disparity suggests that many of the models developed to date are likely to perform better for residents of Asia than for those of Africa and Oceania, given the distinct lifestyles, cultures, and genetic backgrounds of those populations. It is essential that ML for human health be developed and deployed in such a way that it performs equally well on diverse human populations (Huang et al., 2022; Perez-Downes et al., 2024; Starke et al., 2021; Vokinger et al., 2021; Wiens et al., 2019). Reporting study demographics is an important step towards assessing potential biases in models and identifying where additional research efforts are required, and it should be actively encouraged.

Finally, a concerning observation in our corpus of articles was an apparent widespread confusion over how to avoid test set omission and/or data leakage. Such a phenomenon has likely resulted in overestimates of model performance and the predictive value of the microbiome across the field. In fact, for 86% of papers, we were unable to confidently rule out test set omission and/or data leakage based on authors' description of methodology. This finding is in close alignment with Quinn 2021's (Thomas P. Quinn, n.d.) estimate that up to 88% of studies applying ML to microbiomics datasets may contain test set omission and/or data leakage. We infer an urgent need to raise awareness within the microbiomics community about how to avoid and identify model verification failure in microbiomics studies. We refer the reader to Teschendorff, 2019 (Teschendorff, 2019) and Whalen et al., 2022 (Whalen et al., 2022) for extensive discussion of what constitutes data leakage and how it can be avoided.

Acknowledging the limitations of this study is important for contextualizing its contributions and understanding the applicability of its findings in practice. First, we highlight that our review and analysis were based on a sample size of 100 peer-reviewed journal articles. While a larger sample size would have been advantageous, conducting a thorough review of each paper requires significant time and effort, particularly in instances where the methodology is complex, lengthy, and/or vaguely described, and we had to operate within the constraints of resources available to us. Despite the limited sample size, the trends identified in this study likely reflect fundamental issues within the field. We additionally recognize that the selection criteria used to identify papers for inclusion in any review will impact the study's results. Here, we used PubMed to identify peer-reviewed articles with very few constraints, emulating the approach by which a novice reader may gather information. Alternative approaches would potentially yield different core findings, such as focusing on specific ML approaches or technologies, like deep learning, or on specific medical fields with unique characteristics and cultures. Finally, we recognize that our interpretations of the methodology employed by various studies were subjective in some cases, especially where manuscripts provided limited, vague, and/or conflicting descriptions. In order to mitigate the impact of this effect, each study surveyed for this analysis was reviewed by two ML experts. Interpretations of scientific methodology should not be subjective, and this issue highlights the importance of improving transparency and reproducibility in science.

While this work focuses on the application of ML to microbiomics, we expect similar challenges are pervasive across other omics fields, such as genomics, transcriptomics, and metabolomics. For example, genomics research faces analogous issues with widespread data leakage (Whalen et al. 2022; Teschendorff 2019) and underrepresentation of diverse populations in genomic datasets (Sirugo et al. 2019). We highlight two factors that may contribute to why such challenges may be shared across omics disciplines. First, the omics data can be costly, time consuming, and resource intensive to collect and generate. Therefore, many studies work with datasets that are relatively small for ML and/or incorporate data from public datasets (Li et al. 2022). We hypothesize that model verification failure often stems from confusion over how to apply ML best practices when implementing repeated cross-validation procedures in order to mitigate the deleterious effects of small dataset size. Furthermore, investigators may be unaware of how to implement cross-validation schemes that properly account for data being sourced from public databases covering multiple studies. Notably, sources of bias, such as racial disparities in data, stem from broader patterns in medical data collection practices, transcending any one omics discipline (McCradden et al. 2020; Sirugo et al. 2019; Obermeyer et al. 2019; Zou and Schiebinger 2018; Futoma et al. 2020). Second, the application of ML to omics data is highly interdisciplinary. Practitioners typically have a background in either bioinformatics or computer science, each possessing distinct educational foundations. For example, a traditional bioinformatician may not be well-versed in certain machine learning standards, such as the optimal quantity of data required for a specific ML task or the importance of hyperparameter tuning. Conversely, a machine learning expert may lack an understanding of the biological nuances associated with the features used in modeling, such as the advantages of log-ratio transformations in addressing the compositional nature of microbiomics data. Additionally, in the context of medical ML, neither group may be adequately familiar with identifying clinically useful problems or developing solutions that are not only scientifically sound but also ethically responsible and clinically applicable (Wiens et al. 2019). By drawing these parallels, we aim to emphasize that many of the issues and mitigation strategies highlighted here apply broadly across omics research and underscore the potential for cross-disciplinary solutions. In particular, there is significant potential for developing coursework, standards, tools, and frameworks that span multiple omics fields, thereby strengthening a broad spectrum of research.

The development of ML solutions in microbiomics and other areas of healthcare holds substantial potential for positive impact, yet it is also accompanied by numerous ethical concerns pertaining to patient rights, safety, and equity in care. Key considerations include data privacy and security, fairness, transparency and interpretability, and accountability (Char et al. 2020; Zhang and Zhang 2023; Nassar and Kamal 2021). ML is inherently dependent on large datasets, which implies that the methodology used in dataset construction entails significant ethical considerations. Obtaining informed consent from individuals for the use of their data in the development of ML models can be complex, particularly when working with data sourced from large databases that were originally compiled for different purposes (Larson et al. 2020; Iserson 2024; Thapa and Camtepe 2021). Protection of said data is also a major concern, with risks including data breaches, unauthorized access, and misuse of patient data (Thapa and Camtepe 2021). Furthermore, biases in the composition of datasets can adversely affect the model's predictive performance across different demographic groups, potentially exacerbating healthcare disparities if these models are not deployed and made accessible in an equitable manner (McCradden et al. 2020; Obermeyer et al. 2019; Zou and Schiebinger 2018). There are also important considerations around how to ethically apply ML to patients in the clinical setting, particularly concerning transparency, interpretability, and accountability. Many models operate as black boxes, meaning clinicians are unaware of their underlying decision-making process. This lack of visibility poses challenges when applying model recommendations, as clinicians may struggle to identify potential errors or biases within the predictions (Cutillo et al. 2020; Amann et al. 2020). Consequently, questions also arise over accountability. When medical errors arise due to faulty ML predictions, it may be unclear whether responsibility lies with the model developers, the clinician, or the healthcare institution (Naik et al. 2022). As a society, we are still in the process of optimizing approaches to address these ethical questions. It is important for research communities to remain informed and actively engaged in the development of ethical frameworks and guidelines, while regulatory agencies establish comprehensive regulations governing the use of ML in healthcare.

Awareness of widespread deviations from ML best practices within the microbiomics field can be used to improve the quality and reliability of research, in some cases through relatively simple interventions. As ML practitioners, reviewers, and/or regulators, it is our responsibility to optimize the use of scientific resources and ensure that developed ML models are effective, reliable, trustworthy, and fair. Applying best practices to scientific research is essential for expediting the advancement of healthcare and improving patient lives.

**Data availability statement**

Data used in the interactive tutorial was sourced from (Zhu et al., 2020) and is available via the European Nucleotide Archive (ENA) database, accession number ERP111403.

**Acknowledgements**

We thank Viktoria Shalneva (Quantori) for assistance with the graphic design of Figure 5.

**Author declarations**

The authors have no competing interests to declare.

**Works Cited**

Badal, V. D., Vaccariello, E. D., Murray, E. R., Yu, K. E., Knight, R., Jeste, D. V., & Nguyen, T. T. (2020). The gut microbiome, aging, and longevity: a systematic review. *Nutrients*, *12*(12), 3759.

Balle, B., Cherubin, G., & Hayes, J. (2022). Reconstructing training data with informed adversaries. In *2022 IEEE Symposium on Security and Privacy (SP)*. San Francisco, CA, USA: IEEE. http://doi.org/10.1109/sp46214.2022.9833677

Beam, A. L., Manrai, A. K., & Ghassemi, M. (2020). Challenges to the Reproducibility of Machine Learning Models in Health Care. *JAMA: The Journal of the American Medical Association*, *323*(4), 305–306.

Beede, E., Baylor, E., Hersch, F., Iurchenko, A., Wilcox, L., Ruamviboonsuk, P., & Vardoulakis, L. M. (2020). A human-centered evaluation of a deep learning system deployed in clinics for the detection of diabetic retinopathy. In *Proceedings of the 2020 CHI Conference on Human Factors in Computing Systems*. New York, NY, USA: ACM. http://doi.org/10.1145/3313831.3376718

Beleites, C., Neugebauer, U., Bocklitz, T., Krafft, C., & Popp, J. (2013). Sample size planning for classification models. *Analytica Chimica Acta*, *760*, 25–33.

Cadwallader, L., Mac Gabhann, F., Papin, J., & Pitzer, V. E. (2022). Advancing code sharing in the computational biology community. *PLoS Computational Biology*. Public Library of Science San Francisco, CA USA.

Callahan, B. J., DiGiulio, D. B., Goltsman, D. S. A., Sun, C. L., Costello, E. K., Jeganathan, P., … Relman, D. A. (2017). Replication and refinement of a vaginal microbial signature of preterm birth in two racially distinct cohorts of US women. *Proceedings of the National Academy of Sciences of the United States of America*, *114*(37), 9966–9971.

Chan, A.-W., Tetzlaff, J. M., Altman, D. G., Laupacis, A., Gøtzsche, P. C., Krleža-Jerić, K., …

Moher, D. (2013). SPIRIT 2013 statement: defining standard protocol items for clinical trials. *Annals of Internal Medicine*, *158*(3), 200–207.

Chan, A.-W., Tetzlaff, J. M., Gøtzsche, P. C., Altman, D. G., Mann, H., Berlin, J. A., … Moher, D. (2013). SPIRIT 2013 explanation and elaboration: guidance for protocols of clinical trials. *BMJ* , *346*, e7586.

Chen, J., Ryu, E., Hathcock, M., Ballman, K., Chia, N., Olson, J. E., & Nelson, H. (2016). Impact of demographics on human gut microbial diversity in a US Midwest population. *PeerJ*, *4*, e1514.

Collins, G. S., Reitsma, J. B., Altman, D. G., & Moons, K. G. M. (2015). Transparent Reporting of a multivariable prediction model for Individual Prognosis Or Diagnosis (TRIPOD): the TRIPOD Statement. *The British Journal of Surgery*, *102*(3), 148–158.

Cui, H., Wang, C., Maan, H., & Wang, B. (2023, May 1). *scGPT: Towards Building a Foundation Model for Single-Cell Multi-omics Using Generative AI. bioRxiv*. http://doi.org/10.1101/2023.04.30.538439

Curry, K. D., Nute, M. G., & Treangen, T. J. (2021). It takes guts to learn: machine learning techniques for disease detection from the gut microbiome. *Emerging Topics in Life Sciences*, *5*(6), 815–827.

Daliri, E. B.-M., Ofosu, F. K., Chelliah, R., Lee, B. H., & Oh, D.-H. (2021). Challenges and Perspective in Integrated Multi-Omics in Gut Microbiota Studies. *Biomolecules*, *11*(2). http://doi.org/10.3390/biom11020300

Díez López, C., Montiel González, D., Vidaki, A., & Kayser, M. (2022). Prediction of Smoking Habits From Class-Imbalanced Saliva Microbiome Data Using Data Augmentation and Machine Learning. *Frontiers in Microbiology*, *13*, 886201.

Gacesa, R., Kurilshikov, A., Vich Vila, A., Sinha, T., Klaassen, M. A. Y., Bolte, L. A., … Weersma, R. K. (2022). Environmental factors shaping the gut microbiome in a Dutch population. *Nature*, *604*(7907), 732–739.

Gloor, G. B., & Reid, G. (2016). Compositional analysis: a valid approach to analyze microbiome high-throughput sequencing data. *Canadian Journal of Microbiology*, *62*(8), 692–703.

Gloor, G., Macklaim, J. M., Pawlowsky-Glahn, V., & Egozcue, J. (2017). Microbiome datasets are compositional: And this is not optional. *Frontiers in Microbiology*, *8*. http://doi.org/10.3389/fmicb.2017.02224

Gordon-Rodriguez, E., Quinn, T., & Cunningham, J. P. (n.d.). Data Augmentation for Compositional Data: Advancing Predictive Models of the Microbiome. *Advances in Neural Information Processing Systems*.

Graw, S., Chappell, K., Washam, C. L., Gies, A., Bird, J., Robeson, M. S., 2nd, & Byrum, S. D. (2021). Multi-omics data integration considerations and study design for biological systems and disease. *Molecular Omics*, *17*(2), 170–185.

Gregory B. Gloor, Jean M. Macklaim, Vera Pawlosky-Glahn, Juan J. Egozcue. (2017a). Microbiome datasets are compositional: and this is not optional. *Frontiers in Microbiology*, *8*(2224).

Gregory B. Gloor, Jean M. Macklaim, Vera Pawlosky-Glahn, Juan J. Egozcue. (2017b). Microbiome datasets are compositional: and this is not optional. *Frontiers in Microbiology*, *8*(2224).

Haim, N., Vardi, G., Yehudai, G., Shamir, O., & Irani, M. (n.d.). Reconstructing training data from trained neural networks. *Advances in Neural Information Processing Systems*.

Heil, B. J., Hoffman, M. M., Markowetz, F., Lee, S.-I., Greene, C. S., & Hicks, S. C. (2021). Reproducibility standards for machine learning in the life sciences. *Nature Methods*, *18*(10), 1132–1135.

Henderson, P., Islam, R., Bachman, P., Pineau, J., Precup, D., & Meger, D. (2018). Deep reinforcement learning that matters. In *Proceedings of the AAAI conference on artificial intelligence* (Vol. 32).

Heus, P., Damen, J. A. A. G., Pajouheshnia, R., Scholten, R. J. P. M., Reitsma, J. B., Collins, G.

S., … Hooft, L. (2019). Uniformity in measuring adherence to reporting guidelines: the example of TRIPOD for assessing completeness of reporting of prediction model studies. *BMJ Open*, *9*(4), e025611.

Huang, J., Galal, G., Etemadi, M., & Vaidyanathan, M. (2022). Evaluation and mitigation of racial bias in clinical machine learning models: scoping review. *JMIR Medical Informatics*, *10*(5), e36388.

Kaplan, R. C., Wang, Z., Usyk, M., Sotres-Alvarez, D., Daviglus, M. L., Schneiderman, N., … Burk, R. D. (2020). Gut microbiome composition in the Hispanic Community Health Study/Study of Latinos is shaped by geographic relocation, environmental factors, and obesity. *Genome Biology*, *21*(1), 50.

Lakshmanan, V., Robinson, S., & Munn, M. (2020). *Machine Learning Design Patterns*. "O'Reilly Media, Inc."

Lundberg, S. M., & Lee, S.-I. (2017). A unified approach to interpreting model predictions. *Advances in Neural Information Processing Systems*, *30*.

MacEachern, S. J., & Forkert, N. D. (2021). Machine learning for precision medicine. *Genome / National Research Council Canada = Genome / Conseil National de Recherches Canada*, *64*(4), 416–425.

Marcos-Zambrano, L. J., Karaduzovic-Hadziabdic, K., Loncar Turukalo, T., Przymus, P., Trajkovik, V., Aasmets, O., … Truu, J. (2021). Applications of Machine Learning in Human Microbiome Studies: A Review on Feature Selection, Biomarker Identification, Disease Prediction and Treatment. *Frontiers in Microbiology*, *12*, 634511.

McCoubrey, L. E., Elbadawi, M., Orlu, M., Gaisford, S., & Basit, A. W. (2021). Harnessing machine learning for development of microbiome therapeutics. *Gut Microbes*, *13*(1), 1–20.

McCoubrey, L. E., Gaisford, S., Orlu, M., & Basit, A. W. (2022). Predicting drug-microbiome interactions with machine learning. *Biotechnology Advances*, *54*, 107797.

Mirzayi, C., Renson, A., Genomic Standards Consortium, Massive Analysis and Quality Control

Society, Zohra, F., Elsafoury, S., … Waldron, L. (2021). Reporting guidelines for human microbiome research: the STORMS checklist. *Nature Medicine*, *27*(11), 1885–1892.

Moreno-Indias, I., Lahti, L., Nedyalkova, M., Elbere, I., Roshchupkin, G., Adilovic, M., … Claesson, M. J. (2021). Statistical and Machine Learning Techniques in Human Microbiome Studies: Contemporary Challenges and Solutions. *Frontiers in Microbiology*, *12*, 635781.

Mulenga, M., Kareem, S. A., Sabri, A. Q. M., & Seera, M. (2021). Feature extension of gut microbiome data for deep neural network-based colorectal cancer classification. *IEEE*. Retrieved from https://ieeexplore.ieee.org/abstract/document/9319639/

Navas-Molina, J. A., Hyde, E. R., Sanders, J., & Knight, R. (2017). The Microbiome and Big Data. *Current Opinion in Systems Biology*, *4*, 92–96.

NeurIPS 2021 Paper Checklist Guidelines. (2021). Retrieved October 5, 2023, from https://neurips.cc/Conferences/2021/PaperInformation/PaperChecklist

Nushi, B. (2021, February 17). Responsible Machine Learning with Error Analysis. Retrieved August 24, 2024, from https://techcommunity.microsoft.com/t5/ai-machine-learning-blog/responsible-machine-learning-with-error-analysis/ba-p/2141774

Olson, R. S., Cava, W. L., Mustahsan, Z., Varik, A., & Moore, J. H. (2018). Data-driven advice for applying machine learning to bioinformatics problems. In *Biocomputing 2018*. Kohala Coast, Hawaii, USA: WORLD SCIENTIFIC. http://doi.org/10.1142/9789813235533_0018

Papoutsoglou, G., Tarazona, S., Lopes, M. B., Klammsteiner, T., Ibrahimi, E., Eckenberger, J., … Berland, M. (2023). Machine learning approaches in microbiome research: challenges and best practices. *Frontiers in Microbiology*, *14*, 1261889.

Perez-Downes, J. C., Tseng, A. S., McConn, K. A., Elattar, S. M., Sokumbi, O., Sebro, R. A., … Adedinsewo, D. (2024). Mitigating Bias in Clinical Machine Learning Models. *Current Treatment Options in Cardiovascular Medicine*, *26*(3), 29–45.

Pineau, J., Vincent-Lamarre, P., Sinha, K., Larivière, V., Beygelzimer, A., d'Alché-Buc, F., …

Larochelle, H. (2021). Improving reproducibility in machine learning research (a report from the neurips 2019 reproducibility program). *Journal of Machine Learning Research: JMLR*, *22*(1), 7459–7478.

Quinn, T. P. (2021). Stool Studies Don't Pass the Sniff Test: A Systematic Review of Human Gut Microbiome Research Suggests Widespread Misuse of Machine Learning. *arXiv Preprint*.

Quinn, T. P. (n.d.). Stool Studies Don't Pass the Sniff Test: A Systematic Review of Human Gut Microbiome Research Suggests Widespread Misuse of Machine Learning. *arXiv Preprint arXiv:2107. 03611*.

Ravel, J., Gajer, P., Abdo, Z., Schneider, G. M., Koenig, S. S. K., McCulle, S. L., … Forney, L. J. (2011). Vaginal microbiome of reproductive-age women. *Proceedings of the National Academy of Sciences of the United States of America*, *108 Suppl 1*(Suppl 1), 4680–4687.

Reiman, D., & Dai, Y. (2020, May 21). *Using Conditional Generative Adversarial Networks to Boost the Performance of Machine Learning in Microbiome Datasets*. bioRxiv. http://doi.org/10.1101/2020.05.18.102814

Sampling and Splitting. (n.d.). Retrieved October 2023, from https://developers.google.com/machine-learning/data-prep/construct/sampling-splitting/imbalanced-data

Sanders, S., & Giraud-Carrier, C. (2017). Informing the use of hyperparameter optimization through metalearning. In *2017 IEEE International Conference on Data Mining (ICDM)*. New Orleans, LA: IEEE. http://doi.org/10.1109/icdm.2017.137

Sayyari, E., Kawas, B., & Mirarab, S. (2019). TADA: phylogenetic augmentation of microbiome samples enhances phenotype classification. *Bioinformatics* , *35*(14), i31–i40.

Scepanovic, P., Hodel, F., Mondot, S., Partula, V., Byrd, A., Hammer, C., … Milieu Intérieur Consortium. (2019). A comprehensive assessment of demographic, environmental, and host genetic associations with gut microbiome diversity in healthy individuals. *Microbiome*, *7*(1), 130.

Sharma, N. K., Ayyala, R., Deshpande, D., Patel, Y., Munteanu, V., Ciorba, D., … Others. (2024). Analytical code sharing practices in biomedical research. *PeerJ Computer Science*, *10*, e2066.

Shehab, M., Abualigah, L., Shambour, Q., Abu-Hashem, M. A., Shambour, M. K. Y., Alsalibi, A. I., & Gandomi, A. H. (2022). Machine learning in medical applications: A review of state-of-the-art methods. *Computers in Biology and Medicine*, *145*, 105458.

Starke, G., De Clercq, E., & Elger, B. S. (2021). Towards a pragmatist dealing with algorithmic bias in medical machine learning. *Medicine, Health Care and Philosophy*, *24*, 341–349.

Stojnic, R. (2020, April 8). ML Code Completeness Checklist. Retrieved October 5, 2023, from medium.com/paperswithcode/ml-code-completeness-checklist-e9127b168501

Teschendorff, A. E. (2019). Avoiding common pitfalls in machine learning omic data science. *Nature Materials*, *18*(5), 422–427.

Topçuoğlu, B. D., Lesniak, N. A., Ruffin, M. T., 4th, Wiens, J., & Schloss, P. D. (2020). A Framework for Effective Application of Machine Learning to Microbiome-Based Classification Problems. *mBio*, *11*(3). http://doi.org/10.1128/mBio.00434-20

Tsilimigras, M. C. B., & Fodor, A. A. (2016). Compositional data analysis of the microbiome: fundamentals, tools, and challenges. *Annals of Epidemiology*, *26*(5), 330–335.

Vabalas, A., Gowen, E., Poliakoff, E., & Casson, A. J. (2019). Machine learning algorithm validation with a limited sample size. *PloS One*, *14*(11), e0224365.

Varma, S., & Simon, R. (2006). Bias in error estimation when using cross-validation for model selection. *BMC Bioinformatics*, *7*, 91.

Vokinger, K. N., Feuerriegel, S., & Kesselheim, A. S. (2021). Mitigating bias in machine learning for medicine. *Communications Medicine*, *1*(1), 25.

Wang, Q., & Kurz, D. (2022). Reconstructing training data from diverse ML models by ensemble inversion. In *2022 IEEE/CVF Winter Conference on Applications of Computer Vision (WACV)*. Waikoloa, HI, USA: IEEE. http://doi.org/10.1109/wacv51458.2022.00392

Whalen, S., Schreiber, J., Noble, W. S., & Pollard, K. S. (2022). Navigating the pitfalls of applying machine learning in genomics. *Nature Reviews. Genetics*, *23*(3), 169–181.

Wiens, J., Saria, S., Sendak, M., Ghassemi, M., Liu, V. X., Doshi-Velez, F., … Goldenberg, A. (2019). Do no harm: a roadmap for responsible machine learning for health care. *Nature Medicine*, *25*(9), 1337–1340.

Wolff, R. F., Moons, K. G. M., Riley, R. D., Whiting, P. F., Westwood, M., Collins, G. S., … PROBAST Group†. (2019). PROBAST: A Tool to Assess the Risk of Bias and Applicability of Prediction Model Studies. *Annals of Internal Medicine*, *170*(1), 51–58.

Woloszynek, S., Zhao, Z., Chen, J., & Rosen, G. L. (2019). 16S rRNA sequence embeddings: Meaningful numeric feature representations of nucleotide sequences that are convenient for downstream analyses. *PLoS Computational Biology*, *15*(2), e1006721.

Wolpert, D. H., & Macready, W. G. (1997). No free lunch theorems for optimization. *IEEE Transactions on Evolutionary Computation*, *1*(1), 67–82.

Ying, X. (2019). An overview of overfitting and its solutions. In *Journal of physics: Conference series* (Vol. 1168, p. 022022). IOP Publishing.

Zhu, F., Ju, Y., Wang, W., Wang, Q., Guo, R., Ma, Q., … Ma, X. (2020). Metagenome-wide association of gut microbiome features for schizophrenia. *Nature Communications*, *11*(1), 1612.

# Figures

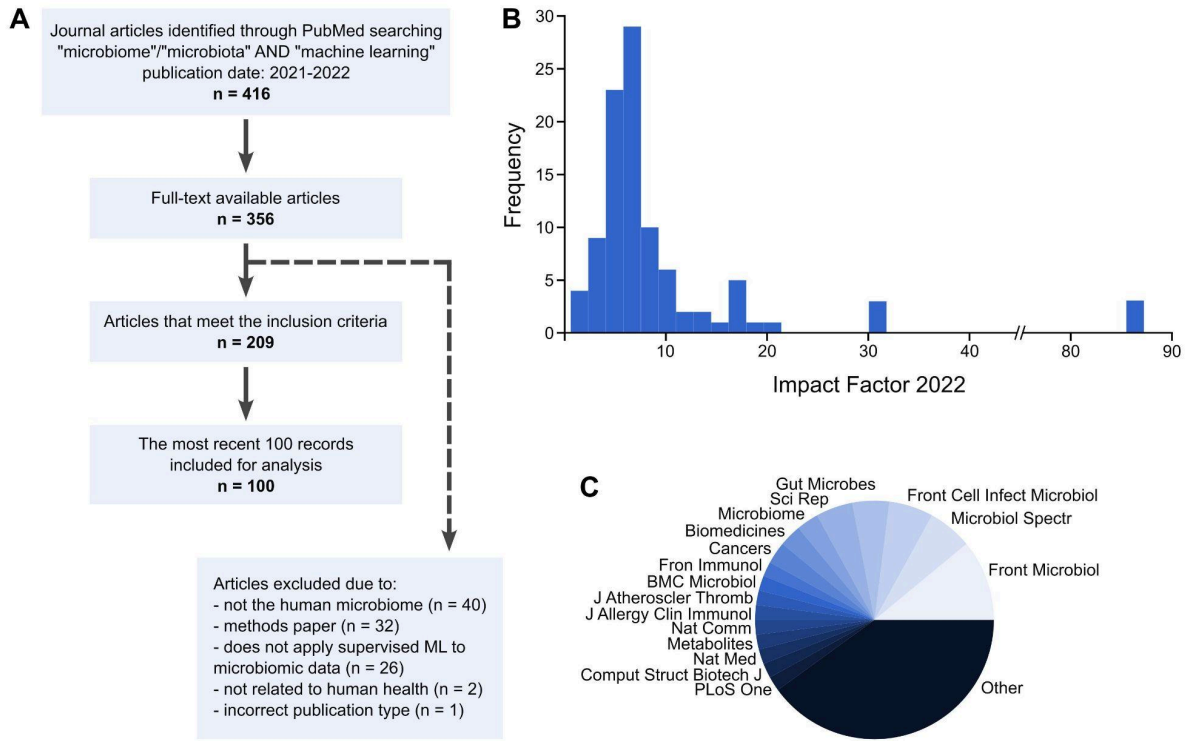

**Figure 1: Overview of studies included for analysis.** To characterize current practices in the application of supervised ML to microbiomics data, we identified 100 peer-reviewed journal articles published in 2021-2022 for in-depth review. **(A)** Flow diagram illustrating the number of records identified, screened, and included in the analysis of current practices. **(B)** Histogram of journal impact factors associated with papers included here (n=100). **(C)** Pie chart showing the number of papers (n=100) per journal. Journals from which two or more studies were sourced have a designated slice, while singleton journals are grouped together under "Other".

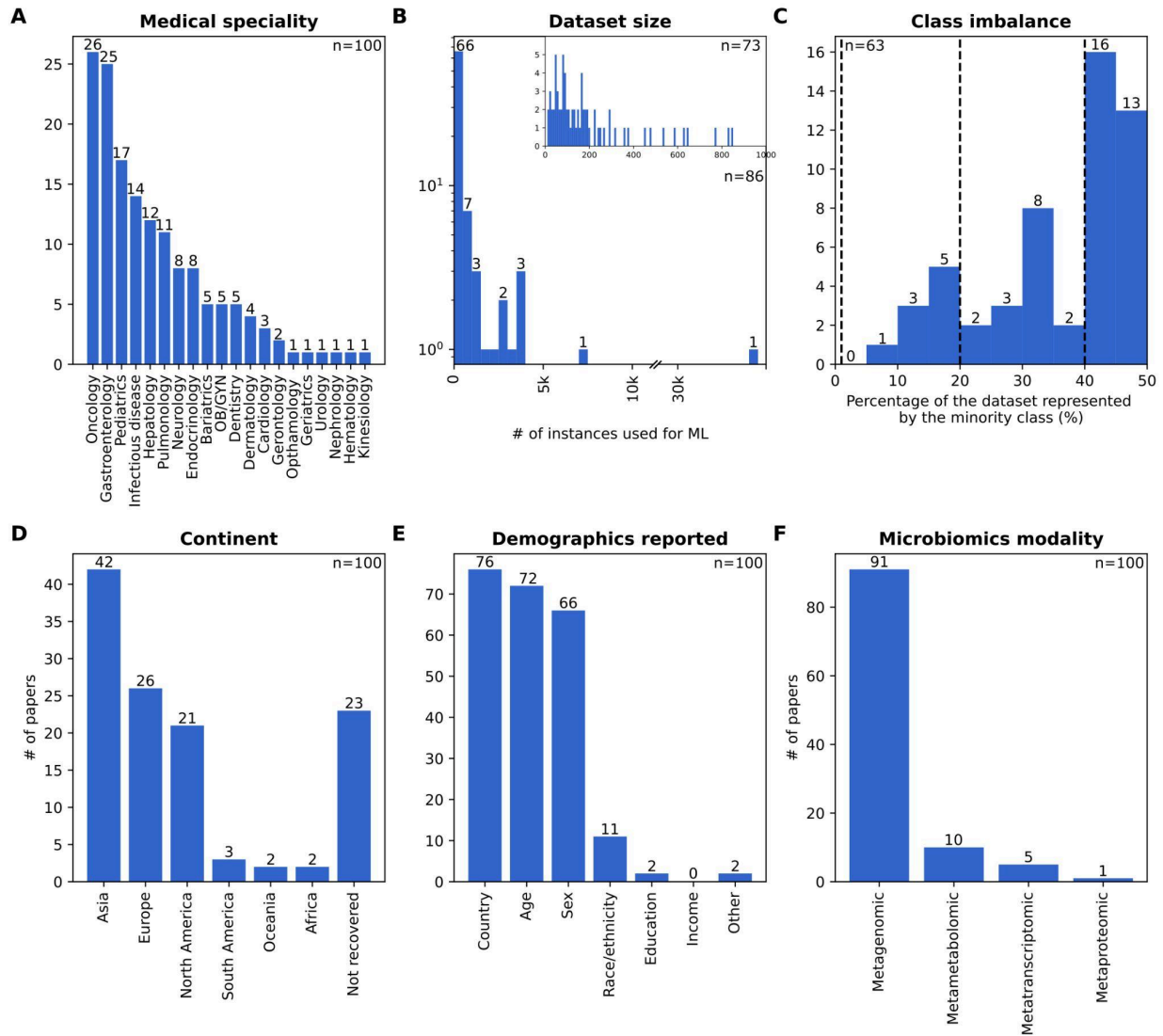

**Figure 2: Properties of microbiome datasets used for ML.** The number of studies considered per analysis is denoted in the top right corner (n=#). Barplots and histograms illustrate: **(A)** Medical specialties or specialities under study, for each of 100 papers. Some studies were categorized under two specialities (e.g., pediatrics and oncology). **(B)** Size of datasets used for ML, for 86 papers from which this information was recovered. **(C)** For 63/85 binary classification tasks for which dataset size per class was recovered, the magnitude of class imbalance between the majority and minority classes. Dashed vertical lines represent, from left to right, the point at which we consider imbalances to be severe, moderate, mild, or not present. **(D)** Geographic distribution of study participant recruitment. Note that in some cases recruitment spanned multiple continents. **(E)** Reporting frequency for the following demographic factors: country of residence, age, sex race/ethnicity, education level, income level / socioeconomic status, other (country of birth, occupation). **(F)** Frequency of microbiomics omics modalities applied. Metagenomic data includes both 16S rRNA gene amplicon and shotgun -based approaches. Note that seven studies used multiple types of omics data to characterize the microbiome.

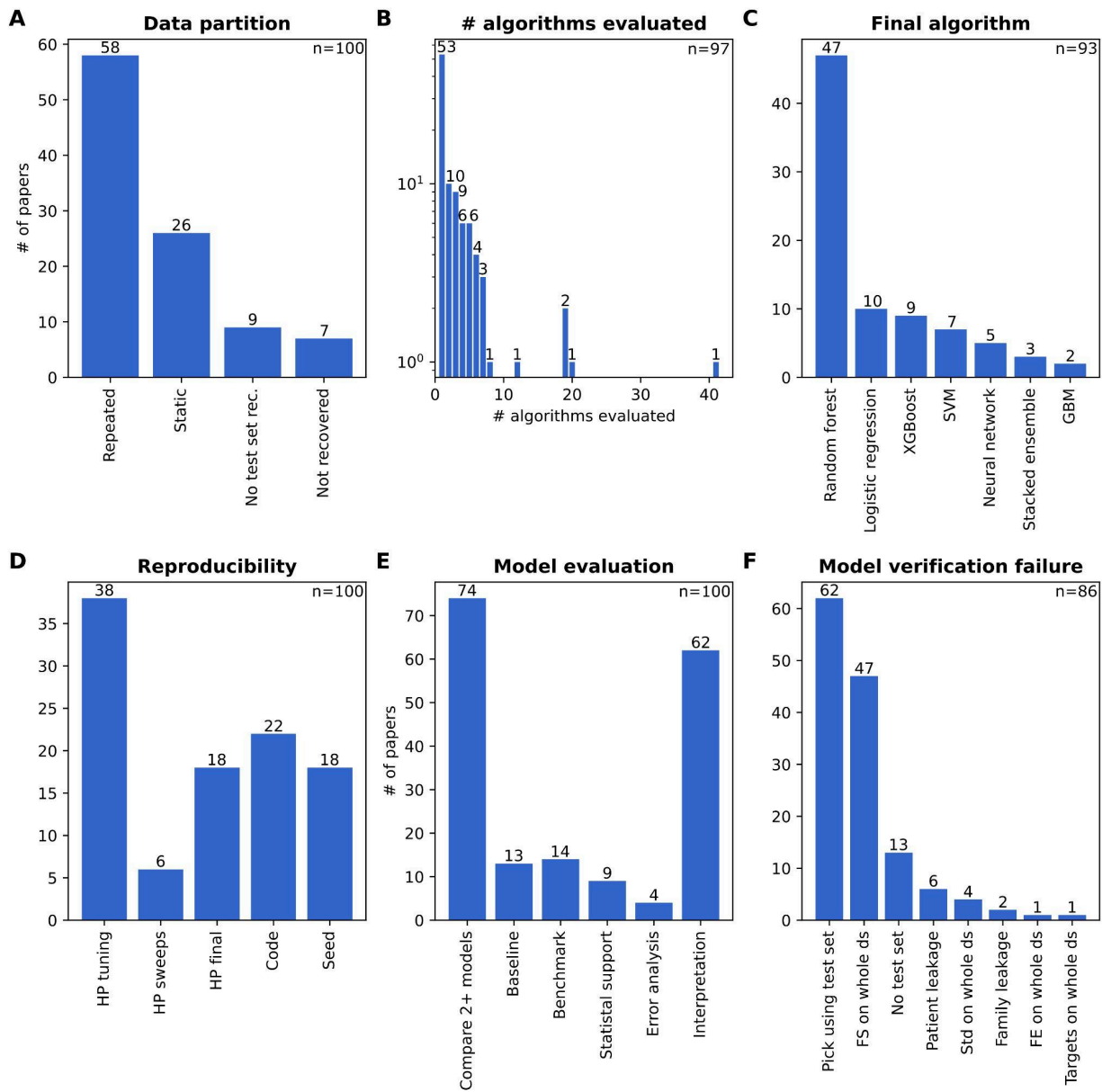

**Figure 3: ML practices in microbiomics.** The number of studies considered per analysis is denoted in the top right corner (n=#). Barplots illustrate: **(A)** Type of data partition (repeated vs static) used for training and validating model performance. Some studies appeared to implement no training-test split or we were not able to recover information on the methodology. **(B)** Number of learning algorithms evaluated per task. **(C)** Learning algorithm used to train the final model, either because it was the only learning algorithm implemented or because it outperformed the alternative/s. **(D)** Subset of factors affecting study reproducibility: the number of papers in which we recovered whether hyperparameter (HP) tuning was performed, the HP sweeps implemented, the final HPs, a functional link to publicly available code, and whether a seed was reportedly set even once for any non-deterministic process related to ML. **(E)** When evaluating the performance the model/s trained, did studies compare two or more models, compare against a baseline, compare against a benchmark, provide statistical support for findings that one model worked better than another, perform error analysis, perform model interpretation. **(F)** Frequency of causes of putative model verification failure. Out of 86 studies containing putative model verification failure, potential causes were as follows: picked final model based on test set results, performed feature selection on the whole dataset (ds), no training-test split / no test set, multiple samples were collected from single study participants but were not restrained to only the training set or only the test set, standardization was performed on the whole dataset, samples were collected from multiple family members but samples from families were not restrained to only the training set or only the test set, feature engineering was performed on the whole dataset, target labels were assigned to samples based on analysis performed on the whole dataset.

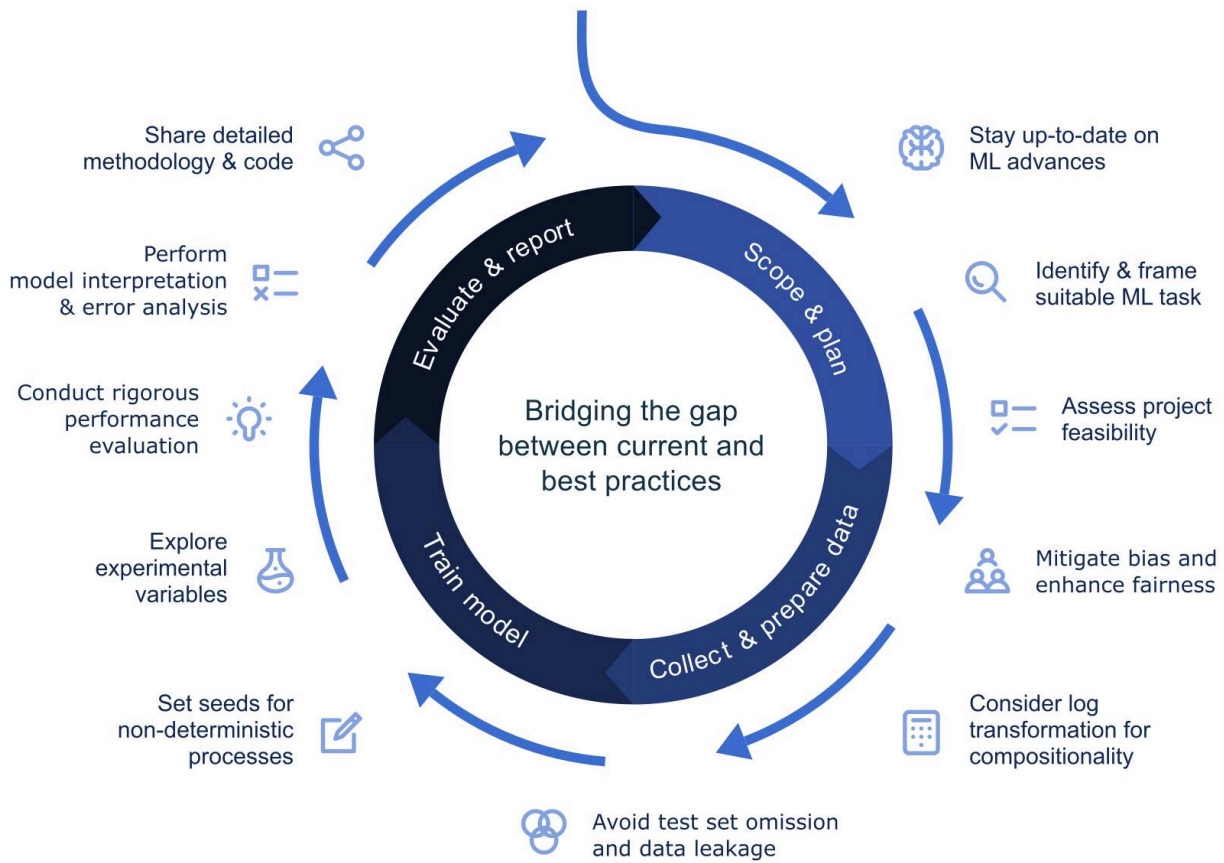

**Figure 4: Bridging the gap between current and best practices in ML for microbiomics.** Action items for how scientists can improve current practices in the application of ML to microbiomics data, thereby accelerating the advancement of science and optimizing the use of scientific resources.

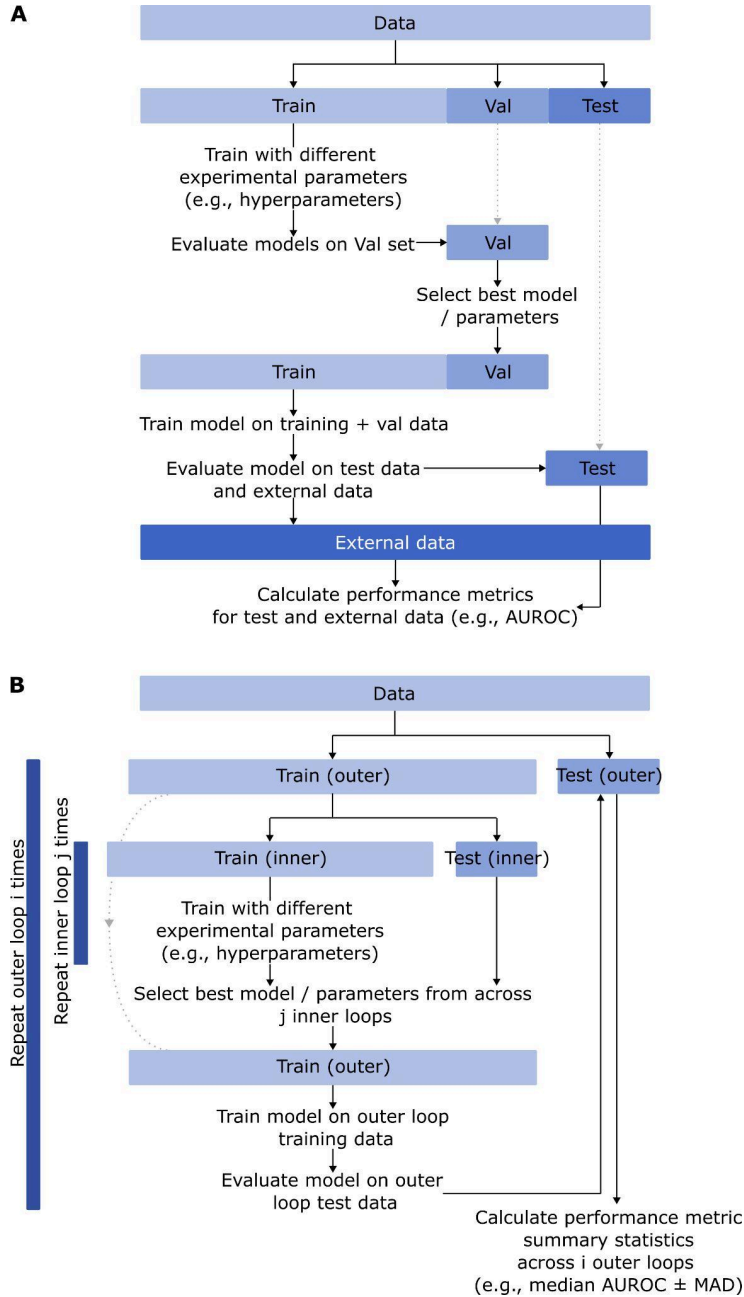

**Figure 5: ML experimental designs that avoid test set omission and leakage. (A)** When dataset size is large, static data partitions (e.g., training, validation, test) sets may be established. Additional validation on an external dataset further supports estimates of model performance and generalizability. When dataset size is small, **(B)** repeated k-fold CV or **(C)** nested CV may be employed to produce unbiased estimates of model performance and/or identifying features whose use is consistently identified as contributing to model performance. For repeated k-fold CV, the dataset is first split into k folds (here, k=5) and each fold is split into a training and set set. This procedure may then be repeated multiple times to increase the number of experimental replicates. Model performance is evaluated on each fold's test set and an aggregate measure of performance is calculated across all folds. For nested CV, data is split

into an outer training and test set *i* times, and a model trained on the training set is evaluated on the test set. Model performance is estimated over *i* models. In order to identify optimal experimental parameters, e.g., hyperparameters or learning algorithm, each training set is split into an inner training and test *j* times. Note that seeds should be set for the splitting process in order to ensure reproducibility. While model validation may be performed on external datasets, if their size is limited estimates may not be robust measures of model performance.

**Supplementary Tables for "Supervised machine learning for microbiomics: bridging the gap between current and best practices"**

Dudek, Natasha Katherine[a,*], Chakhvadze, Mariami[b], Kobakhidze, Saba[b,c], Kantidze, Omar[b], Gankin, Yuriy[a].

**Author affiliations**
[a] Quantori, Cambridge, USA
[b] Quantori, Tbilisi, Georgia
[c] Present address: Free University of Tbilisi, Georgia
*Corresponding author

| **Types of ML tasks** | |
|---|---|
| Medical specialty | Oncology (26), gastroenterology (25), pediatrics (17), infectious disease (14), hepatology (12), pulmonology (11), neurology (8), endocrinology (8), bariatrics (5), OB/GYN (5), dentistry (5), dermatology (4), cardiology (3), gerontology (2), ophthalmology (1), geriatrics (1), urology (1), nephrology (1), hematology (1), kinesiology (1) |
| ML purpose | Disease subtype diagnosis (69), prognosis of disease progression or response to treatment (22), prediction of host characteristic (9) |
| ML approach | Classification (90), regression (10) |
| Number of classes for classification tasks | 2 (85), 3 (3), 4 (1), 9 (1) |
| Biomarker discovery stated as a motivating factor for analysis | Yes (62), no (38) |
| **Dataset characteristics** | |
| Dataset size | ≤100 samples (31), ≤1,000 samples (73), ≤10,000 samples (85), ≤100,000 samples (86), not recovered (14) |
| Dataset size for deep learning | 79 (1), 200 (1), 3131 (1), 3,595 (1), not recovered (1) |
| Data source | Generated by study (68), sourced from external study/ies (28), combination (4) |
| Class imbalance | Balanced (26), mild imbalance (15), moderate imbalance (9), not recovered (37) |
| Continent | Asian (42), Europe (26), North America (21), South America (3), Oceania (2), Africa (2), not recovered (23) |
| Demographics reported | Country (76), age (72), sex (66), race/ethnicity (11), education (2), country of birth (1), occupation (1), income (0) |
| Explicit mention of the term "bias" | Yes (7), no (93) |
| **Generating and selecting features for ML** | |
| Microbiomics modality | Metagenomic (91), metametabolomic (10), |
| | metatranscriptomic (5), metaproteomic (1) |
| ML on multiple microbiome-derived data modalities | Yes (7), no (93) |
| ML on at least one non-microbiome-derived data modalities | Yes (7), no (93) |
| Feature approach | Taxonomic composition (92), gene function (13) |
| If using taxa as features, taxonomic composition | Strain (5), species (32), genus (38), family (2), class (1), combined multiple (8), not recovered (14) |
| If using taxa as features, transformation to account for compositionality | Relative abundance (33-65), centered log-ratio transform (11), other forms of log transform (9), rarefied counts (3), presence-absence (3), cum sum normalization (1), not recovered (8-40) |
| **Training an ML model** | |
| Trained and evaluated ML model across repeated partitionings of the data | Yes (58), no (26), no partition recognized (9), not recovered (7); data partition types included static data partitions, repeated k-fold CV (including LOOCV), Monte Carlo CV, leave-one-dataset/cohort/family-out CV, and nested CV |
| Feature selection | Yes (61), no (31), not recovered (8); feature selection methods included filter methods (e.g., univariate feature selection, DESeq2, LEfSe), wrapper methods (e.g., recursive feature elimination), and embedded methods (e.g., random forest, L1 logistic regression) |
| Compared the performance of at least two ML algorithms prior to selecting a final model | Yes (44), no (56) |
| Best performing ML model when ≥2 models were compared prior to selecting a final model | Random forest (n=47), logistic regression (n=10), XGBoost (n=9), SVMs (n=7), and neural network (n=5), stacked ensemble (3), gradient boosting machine (2) |
| Applied random forest by default without evaluating the performance of other ML algorithms | Yes (33), no (19) |
| Selected a learning algorithm other than random forest after comparing random | Yes (14), no (33) |
| forest's performance against at least one other ML algorithm | |
| Final hyperparameters, at least partially reported | Yes (29), no (71) |
| Hyperparameter search sweeps reported, at least partially | Yes (10), no (90) |
| Provided a functional link to code used to implement experiments described in the study, even if code is only partial | Yes (22), no (75), available upon request (3) |
| Reported setting a seed for any non-deterministic process during the development of an ML model/s | Yes (18), no (82) |
| **Evaluating model performance** | |
| Performed replicates of their ML experiment across multiple data partitions | Yes (58), no (42) |
| If an estimate of model performance was recovered, provided summary statistics characterizing the distribution of results | Yes (33), no (24) |
| If ≥2 models were compared prior to selecting a final model, statistical testing was performed to support claims of model superiority | Yes (11), no (63) |
| Explicitly stated that a baseline or benchmark model was used for evaluating model performance | Yes (10), no (90) |
| Interpretability analysis | Yes (62), no (38); approaches included analysis of SHAP values, Gini impurity decrease index, and permutation feature importance |
| Confident that no test set omission or leakage had occurred | Yes (14), no (86); putative model verification failure occurred due to a) the final model's performance on the test set was reported as an estimate of how well the final model would perform on previously unseen data (62), feature selection was performed on the entire dataset (47), no training-test split appeared to have been implemented (13), multiple samples were collected from single study |

| | participants but were not restrained to only the training set or only the test set (6), standardization was performed on the whole dataset (4), samples were collected from multiple family members but samples from families were not restrained to only the training set or only the test set (2), feature engineering was performed on the whole dataset (1), target labels were assigned to samples based on analysis performed on the whole dataset (1) |
|---|---|

**Table 1: Summary of current practices in the application of ML to microbiomics.** For each practice evaluated, the different approaches in the literature are listed with their frequency shown in brackets (). Note that the entries of rows may not total the number of studies reviewed here (n=100). For example, when summing categories of data used for microbiome studies, some studies employ multiple types of data, and thus the total number of times the different modalities were observed being used is greater than 100.

| Approach | Type of feature | Tools |
|---|---|---|
| **Amplicon sequencing (n=66)** | | |
| Taxonomic (n=64) | OTUs (n=34) | QIIME (n=7), UPARSE (n=6), VSEARCH (n=6), USEARCH (n=5), SortMeRNA (n=2), UCLUST (n=1), SWARM (n=1), Mothur (n=1), not recovered (n=5) |
| | ASVs (n=24) | DADA2 (n=21), UNOISE (n=2), USEARCH (n=1) |
| | ESV (n=1), sOTU (n=1), zOTU (n=1), Other (n=1) | Deblur (n=2), USEARCH (n=1), Kraken (n=1) |
| | Not recovered (n=2) | Not recovered (n=2) |
| Functional (n=3) | Gene and pathway abundances (n=3) | PICRUSt (n=3) |
| **Shotgun sequencing (n=30)** | | |
| Taxonomic (n=28) | Read-based (n=27) | MetaPhlAn (n=12), Centrifuge (n=3), PathSeq (n=3), Kraken / Bracken (n=3), In-house pipeline (n=2), mOTUs (n=2), StrainPhlAn (n=1), VIRGO (n=1), MEGAN (n=1), HIVE-hexagon (n=1) |
| | Genome-resolved (n=2) | Metabat (n=1), In-house pipeline (n=1) |
| Functional (n=10) | Read-based (n=8) | HUMAnN (n=5), NGLess (n=2), UProc (n=1) |
| | Genome-resolved (n=2) | MetaWRAP (n=1), Metabat (n=1) |

**Table 2: Metagenomic entities inferred as and/or used to engineer features for ML.** We loosely use the term "entity" to refer to a microbiome-based feature, such as OTUs, MetaPhlAn-inferred taxa, etc. Of studies surveyed here, 91 studies used metagenomic data for generating ML features, with 14 individual studies having used more than one type of metagenomic entity to do so. For each approach, feature type, and tool used, the number of studies employing it is shown in brackets. In some cases, metagenomic entities were engineered to produce features such as diversity metrics that were used for ML. Note that this table summarizes tools used as they were reported by study authors, e.g., QIIME without specifying which underlying algorithm was used for OTU clustering. Features used by studies for non-ML analyses are not reported here.

**Supplementary Data for "Supervised machine learning for microbiomics: bridging the gap between current and best practices"**

Dudek, Natasha Katherine[a,*], Chakhvadze, Mariami[b], Kobakhidze, Saba[b,c], Kantidze, Omar[b], Gankin, Yuriy[a].

**Author affiliations**
[a] Quantori, Cambridge, USA
[b] Quantori, Tbilisi, Georgia
[c] Present address: Free University of Tbilisi, Georgia
*Corresponding author

**Supplementary Data 1: Journal articles included in the survey of current ML practices for microbiomics.** One hundred peer-reviewed journal articles were surveyed to create a snapshot into current practices in the application of ML to microbiomics data. PubMed Central identifier (PMID), article title, and digital object identified (DOI) are provided below.

1. 35174369, A 'Multiomic' Approach of Saliva Metabolomics, Microbiota, and Serum Biomarkers to Assess the Need of Hospitalization in Coronavirus Disease 2019, 10.1016/j.gastha.2021.12.006
2. 36406449, A multi-omics machine learning framework in predicting the recurrence and metastasis of patients with pancreatic adenocarcinoma, 10.3389/fmicb.2022.1032623
3. 35129058, A specific microbiota signature is associated to various degrees of ulcerative colitis as assessed by a machine learning approach, 10.1080/19490976.2022.2028366
4. 34962914, A systems genomics approach uncovers molecular associates of RSV severity, 10.1371/journal.pcbi.1009617
5. 34997172, Accurate diagnosis of atopic dermatitis by combining transcriptome and microbiota data with supervised machine learning, 10.1038/s41598-021-04373-7
6. 35633725, Alterations of Fungal Microbiota in Patients With Cholecystectomy, 10.3389/fmicb.2022.831947
7. 34787462, Alterations of the Human Lung and Gut Microbiomes in Non-Small Cell Lung Carcinomas and Distant Metastasis, 10.1128/Spectrum.00802-21
8. 36713168, Altered intestinal microbiota composition with epilepsy and concomitant diarrhea and potential indicator biomarkers in infants, 10.3389/fmicb.2022.1081591
9. 35324954, An atlas of robust microbiome associations with phenotypic traits based on large-scale cohorts from two continents, 10.1371/journal.pone.0265756
10. 36232866, Archaea Microbiome Dysregulated Genes and Pathways as Molecular Targets for Lung Adenocarcinoma and Squamous Cell Carcinoma, 10.3390/ijms231911566
11. 35111691, Association Between Vaginal Gardnerella and Tubal Pregnancy in Women With Symptomatic Early Pregnancies in China: A Nested Case-Control Study, 10.3389/fcimb.2021.761153
12. 35536006, Association of Diet and Antimicrobial Resistance in Healthy U.S. Adults, 10.1128/mbio.00101-22
13. 35002989, Bacterial Signatures of Paediatric Respiratory Disease: An Individual Participant Data Meta-Analysis, 10.3389/fmicb.2021.711134
14. 34956135, Can the Salivary Microbiome Predict Cardiovascular Diseases? Lessons Learned From the Qatari Population, 10.3389/fmicb.2021.772736
15. 33612553, Characterization of Salivary Microbiota in Patients with Atherosclerotic Cardiovascular Disease: A Case-Control Study, 10.5551/jat.60608
16. 35701558, Child type 1 diabetes associated with mother vaginal bacteriome and mycobiome, 10.1007/s00430-022-00741-w
17. 35210415, Circulating microbial content in myeloid malignancy patients is associated with disease subtypes and patient outcomes, 10.1038/s41467-022-28678-x
18. 35626941, Clinical Manifestations of Neonatal Hyperbilirubinemia Are Related to Alterations in the Gut Microbiota, 10.3390/children9050764

19. 36266464, Commensal oral microbiota impacts ulcerative oral mucositis clinical course in allogeneic stem cell transplant recipients, 10.1038/s41598-022-21775-3
20. 35228751, Cross-cohort gut microbiome associations with immune checkpoint inhibitor response in advanced melanoma, 10.1038/s41591-022-01695-5
21. 34979898, Determine independent gut microbiota-diseases association by eliminating the effects of human lifestyle factors, 10.1186/s12866-021-02414-9
22. 35318827, Development and evaluation of a colorectal cancer screening method using machine learning-based gut microbiota analysis, 10.1002/cam4.4671
23. 36292203, Diagnosis of Inflammatory Bowel Disease and Colorectal Cancer through Multi-View Stacked Generalization Applied on Gut Microbiome Data, 10.3390/diagnostics12102514
24. 36453905, Different Characteristics in Gut Microbiome between Advanced Adenoma Patients and Colorectal Cancer Patients by Metagenomic Analysis, 10.1128/spectrum.01593-22
25. 36712365, Differential network analysis of oral microbiome metatranscriptomes identifies community scale metabolic restructuring in dental caries, 10.1093/pnasnexus/pgac239
26. 36012106, Discovering Biomarkers for Non-Alcoholic Steatohepatitis Patients with and without Hepatocellular Carcinoma Using Fecal Metaproteomics, 10.3390/ijms23168841
27. 35354069, Early prediction of incident liver disease using conventional risk factors and gut-microbiome-augmented gradient boosting, 10.1016/j.cmet.2022.03.002
28. 36428472, Effect of a Novel E3 Probiotics Formula on the Gut Microbiome in Atopic Dermatitis Patients: A Pilot Study, 10.3390/biomedicines10112904
29. 35562794, Elucidating the role of the gut microbiota in the physiological effects of dietary fiber, 10.1186/s40168-022-01248-5
30. 36098525, Expansion of Opportunistic Enteric Fungal Pathogens and Occurrence of Gut Inflammation in Human Liver Echinococcosis, 10.1128/spectrum.01453-22
31. 35975640, Exploration of Potential Gut Microbiota-Derived Biomarkers to Predict the Success of Fecal Microbiota Transplantation in Ulcerative Colitis: A Prospective Cohort in Korea, 10.5009/gnl210369
32. 36358819, Exploring Gut Microbiome in Predicting the Efficacy of Immunotherapy in Non-Small Cell Lung Cancer, 10.3390/cancers14215401
33. 36357393, Faecal microbiome-based machine learning for multi-class disease diagnosis, 10.1038/s41467-022-34405-3
34. 36428566, Fecal Bacterial Community and Metagenome Function in Asians with Type 2 Diabetes, According to Enterotypes, 10.3390/biomedicines10112998
35. 35903014, Fecal Metabolome and Bacterial Composition in Severe Obesity: Impact of Diet and Bariatric Surgery, 10.1080/19490976.2022.2106102
36. 36289064, Functional profile of host microbiome indicates Clostridioides difficile infection, 10.1080/19490976.2022.2135963
37. 35659906, Gut colonisation by extended-spectrum β-lactamase-producing Escherichia coli and its association with the gut microbiome and metabolome in Dutch adults: a matched case-control study, 10.1016/S2666-5247(22)00037-4
38. 35681218, Gut metabolites predict Clostridioides difficile recurrence, 10.1186/s40168-022-01284-1

39. 34799839, Gut Microbiome Activity Contributes to Prediction of Individual Variation in Glycemic Response in Adults, 10.1007/s13300-021-01174-z
40. 35250836, Gut Microbiota and Targeted Biomarkers Analysis in Patients With Cognitive Impairment, 10.3389/fneur.2022.834403
41. 35173707, Gut Microbiota Composition Is Related to AD Pathology, 10.3389/fimmu.2021.794519
42. 35756062, Gut Microbiota Ecology and Inferred Functions in Children With ASD Compared to Neurotypical Subjects, 10.3389/fmicb.2022.871086
43. 35956358, Gut Microbiota Patterns Predicting Long-Term Weight Loss Success in Individuals with Obesity Undergoing Nonsurgical Therapy, 10.3390/nu14153182
44. 35598439, Higher levels of Bifidobacteria and tumor necrosis factor in children with drug-resistant epilepsy are associated with anti-seizure response to the ketogenic diet, 10.1016/j.ebiom.2022.104061
45. 35040752, Human gut microbiome aging clocks based on taxonomic and functional signatures through multi-view learning, 10.1080/19490976.2021.2025016
46. 36363762, Identification of Human Gut Microbiome Associated with Enterolignan Production, 10.3390/microorganisms10112169
47. 35577971, Identification of shared and disease-specific host gene-microbiome associations across human diseases using multi-omic integration, 10.1038/s41564-022-01121-z
48. 36583106, Identifying distinctive tissue and fecal microbial signatures and the tumor-promoting effects of deoxycholic acid on breast cancer, 10.3389/fcimb.2022.1029905
49. 36032686, Improve the Colorectal Cancer Diagnosis Using Gut Microbiome Data, 10.3389/fmolb.2022.921945
50. 35497193, Inflammatory bowel disease biomarkers of human gut microbiota selected via different feature selection methods, 10.7717/peerj.13205
51. 35923393, Insights of Host Physiological Parameters and Gut Microbiome of Indian Type 2 Diabetic Patients Visualized via Metagenomics and Machine Learning Approaches, 10.3389/fmicb.2022.914124
52. 36382178, Integrated gut microbiome and metabolome analyses identified fecal biomarkers for bowel movement regulation by Bifidobacterium longum BB536 supplementation: A RCT, 10.1016/j.csbj.2022.10.026
53. 35832425, Integrating Choline and Specific Intestinal Microbiota to Classify Type 2 Diabetes in Adults: A Machine Learning Based Metagenomics Study, 10.3389/fendo.2022.906310
54. 35337258, Interpretable prediction of necrotizing enterocolitis from machine learning analysis of premature infant stool microbiota, 10.1186/s12859-022-04618-w
55. 35762814, Leveraging Existing 16SrRNA Microbial Data to Define a Composite Biomarker for Autism Spectrum Disorder, 10.1128/spectrum.00331-22
56. 35050163, Machine Learning Applied to Omics Datasets Predicts Mortality in Patients with Alcoholic Hepatitis, 10.3390/metabo12010041
57. 35495115, Machine learning approach identifies meconium metabolites as potential biomarkers of neonatal hyperbilirubinemia, 10.1016/j.csbj.2022.03.039

58. 35663898, Machine Learning Based Microbiome Signature to Predict Inflammatory Bowel Disease Subtypes, 10.3389/fmicb.2022.872671
59. 36009575, Machine Learning Data Analysis Highlights the Role of Parasutterella and Alloprevotella in Autism Spectrum Disorders, 10.3390/biomedicines10082028
60. 36564713, Machine learning framework for gut microbiome biomarkers discovery and modulation analysis in large-scale obese population, 10.1186/s12864-022-09087-2
61. 36528695, Machine learning-derived gut microbiome signature predicts fatty liver disease in the presence of insulin resistance, 10.1038/s41598-022-26102-4
62. 36180580, Machine-learning algorithms for asthma, COPD, and lung cancer risk assessment using circulating microbial extracellular vesicle data and their application to assess dietary effects, 10.1038/s12276-022-00846-5
63. 36050830, Meta-analysis of caries microbiome studies can improve upon disease prediction outcomes, 10.1111/apm.13272
64. 35920823, Meta-analysis of the microbial biomarkers in the gut-lung crosstalk in COVID-19, community-acquired pneumonia and Clostridium difficile infections, 10.1111/lam.13798
65. 36376984, Microbiome insights into pediatric familial adenomatous polyposis, 10.1186/s13023-022-02569-2
66. 36612118, Microbiome Profiling from Fecal Immunochemical Test Reveals Microbial Signatures with Potential for Colorectal Cancer Screening, 10.3390/cancers15010120
67. 36704050, Mucosal microbiome is predictive of pediatric Crohn's disease across geographic regions in North America, 10.12688/f1000research.108810.2
68. 35677657, Multimodal Data Integration Reveals Mode of Delivery and Snack Consumption Outrank Salivary Microbiome in Association With Caries Outcome in Thai Children, 10.3389/fcimb.2022.881899
69. 35039534, Multimodal deep learning applied to classify healthy and disease states of human microbiome, 10.1038/s41598-022-04773-3
70. 36246273, Oropharyngeal microbiome profiled at admission is predictive of the need for respiratory support among COVID-19 patients, 10.3389/fmicb.2022.1009440
71. 34940616, Physical Activity and Dietary Composition Relate to Differences in Gut Microbial Patterns in a Multi-Ethnic Cohort-The HELIUS Study, 10.3390/metabo11120858
72. 35068612, Pivotal Dominant Bacteria Ratio and Metabolites Related to Healthy Body Index Revealed by Intestinal Microbiome and Metabolomics, 10.1007/s12088-021-00989-5
73. 35875611, Predicting cancer immunotherapy response from gut microbiomes using machine learning models, 10.18632/oncotarget.28252
74. 35983325, Predicting preterm birth through vaginal microbiota, cervical length, and WBC using a machine learning model, 10.3389/fmicb.2022.912853
75. 36539710, Prediction model of poorly differentiated colorectal cancer (CRC) based on gut bacteria, 10.1186/s12866-022-02712-w
76. 35928158, Prediction of Smoking Habits From Class-Imbalanced Saliva Microbiome Data Using Data Augmentation and Machine Learning, 10.3389/fmicb.2022.886201

77. 34427737, Prediction of the occurrence of calcium oxalate kidney stones based on clinical and gut microbiota characteristics, 10.1007/s00345-021-03801-7
78. 35558101, Profiling of the Conjunctival Bacterial Microbiota Reveals the Feasibility of Utilizing a Microbiome-Based Machine Learning Model to Differentially Diagnose Microbial Keratitis and the Core Components of the Conjunctival Bacterial Interaction Network, 10.3389/fcimb.2022.860370
79. 36238647, Recurrence of Early Hepatocellular Carcinoma after Surgery May Be Related to Intestinal Oxidative Stress and the Development of a Predictive Model, 10.1155/2022/7261786
80. 35675435, Risk assessment with gut microbiome and metabolite markers in NAFLD development, 10.1126/scitranslmed.abk0855
81. 36130883, Salivary Microbiota Associated with Peripheral Microvascular Endothelial Dysfunction, 10.5551/jat.63681
82. 34694375, Severe Dysbiosis and Specific Haemophilus and Neisseria Signatures as Hallmarks of the Oropharyngeal Microbiome in Critically Ill Coronavirus Disease 2019 (COVID-19) Patients, 10.1093/cid/ciab902
83. 34428955, SMDI: An Index for Measuring Subgingival Microbial Dysbiosis, 10.1177/00220345211035775
84. 34793867, Stool microbiota are superior to saliva in distinguishing cirrhosis and hepatic encephalopathy using machine learning, 10.1016/j.jhep.2021.11.011
85. 35311577, Surgical Treatment for Colorectal Cancer Partially Restores Gut Microbiome and Metabolome Traits, 10.1128/msystems.00018-22
86. 35436301, The diagnostic value of nasal microbiota and clinical parameters in a multi-parametric prediction model to differentiate bacterial versus viral infections in lower respiratory tract infections, 10.1371/journal.pone.0267140
87. 36587850, The gut microbiome is a significant risk factor for future chronic lung disease, 10.1016/j.jaci.2022.12.810
88. 35854629, The interplay of gut microbiota between donors and recipients determines the efficacy of fecal microbiota transplantation, 10.1080/19490976.2022.2100197
89. 36173308, The Intratumoral Bacterial Metataxonomic Signature of Hepatocellular Carcinoma, 10.1128/spectrum.00983-22
90. 35740540, The Machine-Learning-Mediated Interface of Microbiome and Genetic Risk Stratification in Neuroblastoma Reveals Molecular Pathways Related to Patient Survival, 10.3390/cancers14122874
91. 35873161, The Relationship Between Pediatric Gut Microbiota and SARS-CoV-2 Infection, 10.3389/fcimb.2022.908492
92. 36343772, The upper-airway microbiome as a biomarker of asthma exacerbations despite inhaled corticosteroid treatment, 10.1016/j.jaci.2022.09.041
93. 35928983, The vaginal microbiome is associated with endometrial cancer grade and histology, 10.1158/2767-9764.CRC-22-0075
94. 35013454, Towards a metagenomics machine learning interpretable model for understanding the transition from adenoma to colorectal cancer, 10.1038/s41598-021-04182-y

95. 36452303, Unraveling the impact of Lactobacillus spp. and other urinary microorganisms on the efficacy of mirabegron in female patients with overactive bladder, 10.3389/fcimb.2022.1030315
96. 36574857, Untangling determinants of gut microbiota and tumor immunologic status through a multi-omics approach in colorectal cancer, 10.1016/j.phrs.2022.106633
97. 36483548, Urine metabolomics and microbiome analyses reveal the mechanism of anti-tuberculosis drug-induced liver injury, as assessed for causality using the updated RUCAM: A prospective study, 10.3389/fimmu.2022.1002126
98. 35017168, Using Machine Learning to Identify Metabolomic Signatures of Pediatric Chronic Kidney Disease Etiology, 10.1681/ASN.2021040538
99. 35311570, Vaginal Atopobium is Associated with Spontaneous Abortion in the First Trimester: a Prospective Cohort Study in China, 10.1128/spectrum.02039-21
100. 36109637, Variability of strain engraftment and predictability of microbiome composition after fecal microbiota transplantation across different diseases, 10.1038/s41591-022-01964-3